\documentclass[12pt]{article}
\usepackage{latexsym} \usepackage{epsf}
\usepackage{epsfig}
\usepackage{a4}
\usepackage{amsfonts}
\usepackage[colorlinks]{hyperref}
\usepackage{color}
\newcommand{\ba}{\begin{array}}
\newcommand{\ea}{\end{array}}
\newcommand{\be}{\begin{equation}}
\newcommand{\ee}{\end{equation}}
\newcommand{\nn}{\nonumber}
\newcommand{\bea}{\begin{eqnarray}}
\newcommand{\ena}{\end{eqnarray}}
\newcommand{\beas}{\begin{eqnarray*}}
\newcommand{\enas}{\end{eqnarray*}}

\renewcommand{\theequation}{\thesection.\arabic{equation}}

\begin{document}

\begin{center}

{\Large{New %
series of multi-parametric solutions to GYBE: quantum gates and integrability
}}
\end{center}

\begin{center}{\bf
Shahane A. Khachatryan \footnote{e-mail:{\sl shah@mail.yerphi.am}
 }}
 \\ A.~I.~Alikhanyan National Science Laboratory (YerPhI), 
 \\ Alikhanian Br. str. 2, Yerevan 36, Armenia
\end{center}

\begin{abstract}
We obtain  two
 series of spectral parameter dependent solutions to the generalized Yang-Baxter equations (GYBE), for definite types of $N_1^2\times N_2^2$ matrices with general dimensions $N_1$ and $N_2$. Appropriate extensions are presented  for the inhomogeneous GYBEs. The first series of the solutions includes as particular cases the $X$-shaped trigonometric braiding matrices. For construction of the second series the colored and graded permutation operators are defined, and multi-spectral parameter Yang-Baxterization is performed. For some examples the corresponding integrable models are discussed. The unitary solutions existing in these two series  can be considered as generalizations of the multipartite Bell matrices in the quantum information theory.
\end{abstract}

{\small\tableofcontents}


\section{Introduction}


The area of the quantum exactly solvable models in recent decades admitted close relationship with the disciplines of quantum computing. The generalized Yang-Baxter equations (GYBE) are investigated intensively for depicting unitary R-matrices on the braid group representations, as candidates for complex quantum gates, used in the quantum information processing. Yang-Baxter equations (YBE), initiating as star-triangle equations for solving  concrete models in statistical mechanics, have acquired an important role in solvability of the many-body systems, being key equations in the theory of quantum  integrability \cite{Yang}-\cite{PAY}. Later YBE found applications in quantum field theory, quantum topology, string theory, quantum groups.   And there are emerged numerous works (\cite{FKL}-\cite{YG}), revealing the importance of YBE in the  quantum information theory. Here the information is encoded in quantum states and processed by unitary evolution operators. The quantum computation process is modeled by quantum circuits, which are the sequences of the quantum gates, the quantum analogs of classical logic gates. The n-qubit quantum gates are described by  unitary matrices defined on the superpositions of the tensor products of $n$ states-qubits (the basic unit of the quantum information, quantum two-dimensional analog of the classical binary information unit - bit). As the logical gates are basic blocks, it is natural to analyze the cases with small n-s, $n=1, 2, 3$ \cite{PAY,FKL,Kit}.
 For $n=2$, particularly, the quantum gates can be associated with two-state unitary  $\check{R}$-matrices which are the solutions of the YBE, relying on the revealed topological  properties of the quantum gates - braid group symmetries \cite{art}-\cite{LKSL}.  In topological quantum computation, the unitary representations of the braid group can be described by braiding anyons  \cite{RZWG,GHR}. The $p^2 \times p^2$ (qupit \cite{qupit}) generalizations of the quantum gates as braid group representations and corresponding YBE solutions are discussed in \cite{NSSFS,RZWG,WSZX} associated with the $Z_p$ parafermion.  Generalized YBE, with $\check{R}$-matrices acting on the tensor product of more than two vector states, are developed in \cite{RZWG,Chen,part}, showing, that braiding unitary solutions to GYBE  can be used in quantum information processing.

 In this paper we consider spectral parameter dependent solutions to inhomogeneous GYBE  for arbitrary dimensional matrices of two types, which both are the extensions of $4\times 4$ eight-vertex trigonometric solution: extension by form (the first kind  - X-shaped) and by construction (the second kind). In the work \cite{SHKHAS} entire description is done of the multi-parametric solutions of YBE for the simplest case of  $4 \times 4$ matrices (see also \cite{Baxter,SHKH113,SHKH14}). Here we obtain direct generalizations of all these solutions for even dimensional $X$-shaped \cite{RZWG}  matrices.  In this paper we  present a thorough investigation for even dimensional matrices, the odd dimensional cases we describe comparably in a brief way: the main results are obtained and presented, and the detailed analysis  with some peculiar solutions will be presented in the further work \cite{SHKHoi}.

 In Section 2 the GYB equations are analysed in the context of the integrability of 1D quantums models. The inhomogeneous extensions of GYBE are proposed, and in Subsection 2.1 we emphasize the important differences of these cases in comparison to inhomogeneous YBE. We have discussed the inhomogeneous $N_1^2\times N_2^2$ dimensional  ``check'' or braiding $\check{R}$-matrices defined on the space $V_{N_1}\otimes V_{N_2}$.  The multi-parametric extensions of the known solutions to GYBE are constructed in Section 3. The general  even dimensional $X$-shaped solutions to GYBE, their parameterizations and the corresponding integrable models are discussed, the particular examples (for the solutions to inhomogeneous $(2,3,1)$-GYBE) are presented  in Appendix.   The peculiarities of the odd-dimensional solutions we discuss in Subsection 3.1. Then in Section 4 we sketch also 
 the new perspectives given by the obtained $R$-matrices in the quantum information theory. In Section 5 the graded and colored representations of the braid group are considered, with corresponding Yang-Baxterization, bringing to trigonometric solutions. After appropriate re-parametrization they can considered as new second kind solutions to YBE \cite{ZhLKL}. The examples of $16\times 16$ solutions are discussed in details in Subsections 5.2 and 5.3.

\section{YBE and generalized YBE.}
 \addtocounter{section}{0}\setcounter{equation}{0}

 The  generalized  Yang-Baxter equations
can be defined by the following relations, as in \cite{RZWG,Chen}
\bea\label{GYBE}
(\check{R}\otimes I^p) (I^p\otimes \check{R}) (\check{R}\otimes I^p)=(I^p\otimes \check{R}) (\check{R}\otimes I^p)(I^p\otimes \check{R}),
\ena
Here the matrix $\check{R}$ is defined on the product $\bigotimes^k V$, the unity operator $I^p$ acts as an identity operator on the product of the
vector states $\bigotimes^p V$. These equations are called GYBEs of type $(d,k,p)$, where $d=Dim[V]$.  Far distanced $R$-matrices, which have no overlap of the acting states, are commutative. Thus, these equations reproduce the braid
algebra \cite{art,jones}, representing $b_i=I\otimes...\check{R}\otimes I^p...\otimes I$,
$b_{i+1}=I\otimes...I^p\otimes\check{R}...\otimes I$,
\bea
b_i b_{i+1} b_i=b_{i+1} b_i b_{i+1},\qquad b_i b_j=b_j b_i, \quad |i-j|>1.\label{bb}
\ena
 Here there is used the ``check'' notation for the matrices, since in the particular case of
 $(d,2,1)$ these equations  coincide with the ordinary YBE \cite{Tu} in check formulation. This is true for the homogeneous cases, when the states on which the $\check{R}$-matrix acts, are identical.%

  Before proceed further let us  consider the  generalization of the inhomogeneous  YBE  to  the  inhomogeneous GYBE at least by the following two ways.
   \paragraph{i. The  inhomogeneous GYBE - $(\{d\},\{k,k'\},\{p,p'\})$, $k+p=k'+p'$} can be presented in the following way:
\bea\label{GYBEi}
(\check{R}\otimes I^p) (I^{p'}\otimes \check{R}') (\check{R}\otimes I^p)=
(I^{p'}\otimes \check{R}') (\check{R}\otimes I^p)(I^p\otimes \check{R}),
\ena
  Here the system of the equations acts on the product  of the identical spaces $V$, with the same dimension $d$, and the matrix $R'$ acts on the product $\bigotimes^{k'}V$, so that  $[k+p=p'+k']$. At $k=k'$ this system  coincides with the ordinary GYBE.  Let us present the generalisations of the braiding relations to inhomogeneous or alternating cases: in (\ref{bb}) one can take the following realization, $b_{2i}=I\otimes...\check{R}\otimes I^p...\otimes I$,
  ${b}_{2i+1}=I\otimes...I^{p'}\otimes{\check{R}}'...\otimes I$.

    \paragraph{ii. The full  inhomogeneous GYBE - $(\{d_1,...,d_{k+p}\},\{k,k'\},\{p,p'\})$, $k+p=k'+p'$}.

   Let us extend definition of the action space to the inhomogeneous product of the
   vector states $V_i$ of number $[k+p](=[k'+p'])$ with different characteristics of the vector states (in general with different dimensions $d_i$), and intend the following inhomogeneous matrices $\check{R}:V_1\otimes...\otimes V_k$ and $\check{R}':V_1\otimes...\otimes V_{k'}$; in the  discussion below we use  $\check{R}_{i^1...i^n}$,
   where the indexes $1,...,n$ show the position of the vector space, while the
   index $i^n$ denotes the nature of the vector space
   %
%
\bea
\left(\check{R}_{i^1...i^k}\otimes I_{i^{k+1}}\otimes....I_{i^{p+k}}\right)\left(I_{i^{1}}\otimes....I_{i^{p'}}\otimes \check{R}_{i^{p'+1}...i^{k'+p'}}\right)\times\\
\left(\check{R}_{i^1...i^k}\otimes I_{i^{k+1}}\otimes....I_{i^{p+k}}\right)=\left(I_{i^{1}}\otimes....I_{i^{p'}}\otimes \check{R}_{i^{p'+1}...i^{k'+p'}}\right)\times\nn\\
\left(\check{R}_{i^1...i^k}\otimes I_{i^{k+1}}\otimes....I_{i^{p+k}}\right)\left(I_{i^{1}}\otimes....I_{i^{p'}}\otimes \check{R}_{i^{p'+1}...i^{k'+p'}}\right).\nn
\ena

 The unitary solutions to GYBE, constituting the braid group representations, are connecting  to the quantum computing theory  \cite{LKSL}.
  However, even for the inhomogeneous case, when at general $d_i\neq d_j$, the GYBE solutions  can also be used to the construction  of the integrable models, as we describe below.
The spectral parameter dependent versions of GYBE are defined just as for the case of YBE, thus ensuring the connection with the integrable models:
\bea\label{GYBEsp}
(\check{R}(u,v)\otimes I^p) (I^p\otimes \check{R}(u)) (\check{R}(v)\otimes I^p)=(I^p\otimes \check{R}(v)) (\check{R}(u)\otimes I^p)(I^p\otimes \check{R}(u,v)).
\ena
\subsection{The differences between the usual inhomogeneous YBE and the GYBE}

Let us consider a graphical interpretation  of GYBE, $(d,k,p)$, which act on the product of the states $\bigotimes^{k+p}V_{d}$, 
(Fig. 1), with notation $\check{R}^k$ for the matrix acting on the product of the states: $\bigotimes^k V_{d}$.
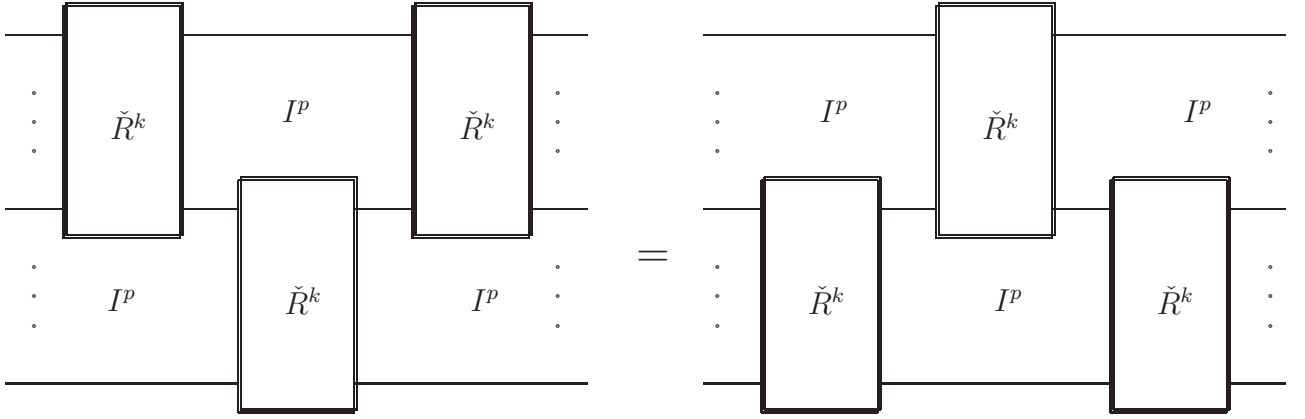
\begin{figure}[t]
\unitlength=11pt
\begin{picture}(100,17)(0,1)
\multiput(0,2)(0,4){1}{\line(1,0){8}}
\multiput(12,2)(0,4){1}{\line(1,0){8}}
\multiput(0,8)(6,0){4}{\line(1,0){2}}
\multiput(0,14)(0,2){1}{\line(1,0){2}}
\multiput(6,14)(0,2){1}{\line(1,0){8}}
\multiput(18,14)(0,2){1}{\line(1,0){2}}
\put(21.7,6){\Large{$=$}}
\multiput(24,14)(0,2){1}{\line(1,0){8}}
\multiput(36,14)(0,2){1}{\line(1,0){8}}
\multiput(24,8)(6,0){4}{\line(1,0){2}}
\multiput(24,2)(0,2){1}{\line(1,0){2}}
\multiput(30,2)(0,2){1}{\line(1,0){8}}
\multiput(42,2)(0,2){1}{\line(1,0){2}}

\newsavebox{\rmatrx}

\sbox{\rmatrx}{\begin{picture}(0,0)(4,8)
\multiput(0,0)(4,0){2}{\line(0,1){8}}
\multiput(0,0)(0,8){2}{\line(1,0){4}}
\put(1.6,3.4){$\check{R}^k$}
\end{picture}}

\newsavebox{\rmatrxk}

\sbox{\rmatrxk}{\begin{picture}(0,0)(4,8)
\multiput(0,0)(4,0){2}{\line(0,1){8}}
\multiput(0,0)(0,8){2}{\line(1,0){4}}
\end{picture}}

\multiput(6,15)(12,0){2}{\usebox{\rmatrx}}
\multiput(6.1,15.1)(12,0){2}{\usebox{\rmatrxk}}
\put(12,9){\usebox{\rmatrx}}
\put(12.1,9.1){\usebox{\rmatrxk}}
\multiput(3.5,4.5)(12.5,0){2}{$I^p$}\put(9.5,11){$I^p$}
\multiput(1,4)(0,1){3}{\circle{0.1}}
\multiput(1,10)(0,1){3}{\circle{0.1}}
\multiput(19,4)(0,1){3}{\circle{0.1}}
\multiput(19,10)(0,1){3}{\circle{0.1}}

\multiput(24.5,4)(0,1){3}{\circle{0.1}}
\multiput(24.5,10)(0,1){3}{\circle{0.1}}
\multiput(43.5,4)(0,1){3}{\circle{0.1}}
\multiput(43.5,10)(0,1){3}{\circle{0.1}}

\multiput(30,9)(12,0){2}{\usebox{\rmatrx}}
\multiput(30.1,9.1)(12,0){2}{\usebox{\rmatrxk}}
\put(36,15){\usebox{\rmatrx}}
\put(36.1,15.1){\usebox{\rmatrxk}}
\multiput(28,11)(12.5,0){2}{$I^p$}\put(34,4.5){$I^p$}

\end{picture}
\caption{GYBE of type $\{d,k,p\}$}
\end{figure}

Note, that the equations $\{d,2p,p\}$  can also be considered as usual YBE, which act on the product of the states  $V'\otimes V'\otimes V'$, with $V'=\bigotimes^p V$. From this point of view, the other cases $k\neq 2p$,
can be considered as the equations on the space $V'\otimes V{''}\otimes V'$, with $V'=\bigotimes^p V$, $V{''}=\bigotimes^{k-p} V$.
\bea \label{YBEII}
[\check{R}_{V' V{''}}\otimes I_{V'}][I_{V'}\otimes \check{R}_{V{''}V'} ][\check{R}_{V' V{''}}\otimes I_{V'}]=[I_{V'}\otimes \check{R}_{V{''}V'}][\check{R}_{V'V{''}}\otimes I_{V'}][ I_{V'}\otimes\check{R}_{V{'}V{''}}].
\ena
In the Fig.1 the way of this kind of the combination of the vector spaces into two big spaces, is apparent.   We see that these equations are quite different  from the usual inhomogeneous YBE defined on the spaces $V'\otimes V{''}\otimes V'$  in the ``check''-matrix formulation \cite{GSA}:
\bea
[\check{R}_{V'V{''}}\otimes I_{V'}][I_{V''}\otimes \check{R}_{V{'}V'}][\check{R}_{V{''}V{'}}\otimes I_{V'}]=[ I_{V'}\otimes \check{R}_{V{''}V'}][\check{R}_{V'V{'}}\otimes I_{V''}][ I_{V'}\otimes \check{R}_{V{'}V{''}}].\label{YBEI}
\ena
This is the usual braiding form of  ordinary YB  equations, the ``check''-operation do not affair the form of the
equations, it only permutes the upper indexes of the matrices in the same YBEs. In the Fig. 2 there are shown two type of the matrices $R_{V'\otimes V{''}}$, acting on the space $\bigotimes^k V$, which appear  in the equations (\ref{YBEII}) and (\ref{YBEI}).

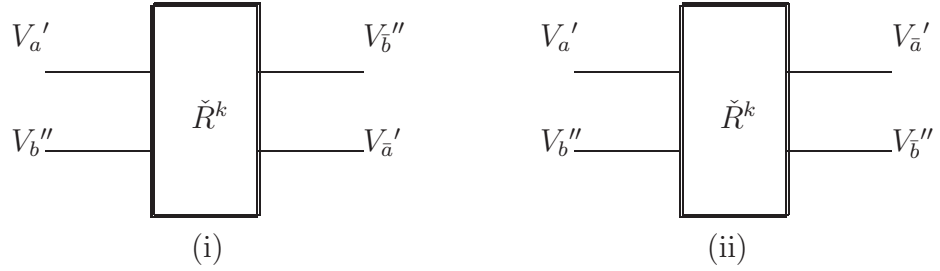
\begin{figure}[h]
\unitlength=10pt
\begin{picture}(100,10)(-1.7,1)
\newsavebox{\rmatrxt}

\sbox{\rmatrxt}{\begin{picture}(0,0)(4,8)
\multiput(0,0)(4,0){2}{\line(0,1){8}}
\multiput(0,0)(0,8){2}{\line(1,0){4}}
\put(1.5,3.4){$\check{R}^{k}$}
\end{picture}}

\newsavebox{\rmatrxkt}

\sbox{\rmatrxkt}{\begin{picture}(0,0)(4,8)
\multiput(0,0)(4,0){2}{\line(0,1){8}}
\multiput(0,0)(0,8){2}{\line(1,0){4}}
\end{picture}}
\multiput(16,9)(20,0){2}{\usebox{\rmatrxt}}
\multiput(16.1,9.1)(20,0){2}{\usebox{\rmatrxkt}}

\multiput(8,3.5)(0,3){2}{\line(1,0){4}}
\multiput(16,3.5)(0,3){2}{\line(1,0){4}}
\put(6.7,3.5){$V_b{''}$}
\put(6.7,7.5){$V_a{'}$}
\put(20,3.5){$V_{\bar{a}}{'}$}
\put(20,7.5){$V_{\bar{b}}{''}$}
\put(13.5,-0.5){(i)}

\multiput(28,3.5)(0,3){2}{\line(1,0){4}}
\multiput(36,3.5)(0,3){2}{\line(1,0){4}}
\put(26.7,3.5){$V_b{''}$}
\put(26.7,7.5){$V_a{'}$}
\put(40,3.5){$V_{\bar{b}}{''}$}
\put(40,7.5){$V_{\bar{a}}{'}$}
\put(33,-0.5){(ii)}

\end{picture}
\caption{The usual $\check{R}_{ab}^{\bar{b}\bar{a}}$(${R}_{ab}^{\bar{a}\bar{b}}$)-matrix in YBE (i), $\check{R}_{ab}^{\bar{a}\bar{b}}$-matrix in GYBE (ii)}
\end{figure}
 If in (\ref{YBEII}) the $\check{R}$-matrices are acting only on the space $\bigotimes^{k}V$, for the case (\ref{YBEI}) there are also matrices defined on the space $\bigotimes^{2p}V$.
  Recall, that the non-check $R$-matrix acts as $P \check{R}$, so in the Fig. 2 (i) it appears with the permutation of the upper indexes: $R_{ab}^{\bar{a}\bar{b}}$. We see, that the given difference between the usual inhomogeneous YBE and GYBE is encoded in the definition of the ``check''-matrices. For the first case it acts as the usual braiding matrix, Fig. 2 (i). In the second case, correspondingly, the graphical representation of the ``check''-matrix is presented in the Fig. 2 (ii), and can be considered in some sense as ``reflecting'' matrix. However they both satisfy the braiding relations. 

 For the inhomogeneous equations, the vector states, on which the $R$-matrices act, can differ not only by their dimensions, also by ``inner'' characteristics.  As example, the intertwiner $R$-matrices, defined
  in the theory of the quantum groups, satisfy to the ordinary YBE \cite{GSA}, and in the ``check'' formulation commute with the generators of the quantum group. There the vector states $V$ constitute the representations of the quantum groups and can differ one from other  by the characteristics of the  representations - the eigenvalues of the center of the group \cite{GSA,SHKH113}. An
interesting example is the case of two-dimensional cyclic states of the $sl_q(2)$ group at $q^4$ \cite{SHKH113}, where two vector spaces are of the same dimension, but have different characteristics, which  bring to various kind of invariant solutions to YBE. And one can find  how to operate with the inhomogeneous $R$-matrices in ``check'' formulation for the usual Yang-Baxter equations, particularly in the works \cite{SHKKH09,SHKH113}, where  we have presented  the solutions with the quantum group
 symmetry including the permuted projection operators between the representation states.

\paragraph{The case $p\geq k$.} In this section we have considered the non-trivial case, when $p<k$. However, one can consider also the remaining cases. As it can be easily followed from the presented equations and the Fig. 1, at  $p \geq k$ the generalised YB equations factorize  into two similar operator equations. The operators $[\check{R}\otimes I^p]$ and $[I^p\otimes \check{R}]$ are commutative and act non-trivially on different, non-intersecting spaces: $V_{1}\otimes...\otimes V_{k}$ and $V_{p+1}\otimes...\otimes V_{p+k}$ (we use the notations introduced within the item (ii)). 
 The right-hand  and left-hand sides of GYBE act as unity operators on the vector spaces $V_{k+1}\otimes...\otimes V_{p}$.   Meanwhile on the product $V_{1}\otimes...\otimes V_{k}$ (or $V_{p+1}\otimes...\otimes V_{p+k}$) the spectral parameter dependent GYBEs (\ref{GYBEsp}) in the cases $p \geq k$  take the following matrix product form, with an arbitrary function $F(u,v)$:
\bea
\sum_{k,p}[{\check{R}}(u,v)]_{ij}^{kp} [{\check{R}}(v)]_{kp}^{i'j'}=F(u,v)[{\check{R}}(u)]_{ij}^{i'j'}.
\ena
For the particular case of $\{2,2,p\}$, $p\geq 2$, the possible solutions (besides of the ones equivalent to unity matrices) are the following ones, with independent constants $\{\theta\epsilon,\; t,q,\; \theta_i\}$ and a function $\bar{F}(u)\neq 0$, $F(u,v)=\frac{\bar{F}(u-v)\bar{F}(v)}{\bar{F}(u)}$, and with the difference property of the
spectral parameters in GYBE $(u,v)\Rightarrow(u-v)$:
\bea\label{r22qt}\frac{\check{R}_e(u)}{\bar{F}(u)}=\left(\!\ba{cccc}
e^{ \theta_1 u}&0&0&0\\
0&e^{\theta_2 u}&0&0\\
0&0&e^{\theta_3 u}&0\\
0&0&0&e^{\theta_4 u}\ea\!\right),\;
\frac{\check{R}(u)}{\bar{F}(u)}=\left(\!\ba{cccc}
\cos[\theta u]&0&0&q\sin[\theta u]\\
0&\cos[\epsilon u]&t \sin[ \epsilon u]&0\\
0&\frac{-\sin[\epsilon u]}{t}&\cos[ \epsilon u]&0\\
\frac{-\sin[\theta u]}{q}&0&0&\cos[\theta u]\!\ea\right)
\ena
When $t=\pm 1,\epsilon=\theta$, $q=e^{i \phi}$, $\bar{F}(u)=e^{i \gamma u}$ and $\bar{u}=\tan[\theta u]$, the second  matrix  is the well known unitary trigonometric solution to YBE with $(\bar{u},\bar{v})\Rightarrow(\frac{\bar{u}-\bar{v}}{1-\bar{u} \bar{v}})$ (the second kind solution to YBE \cite{ZhLKL}, with relativistic rule of the summation of the spectral parameters - ``velocities''),
\bea\label{gYBEss}
\check{R}_{ij}(\frac{\bar{u}-\bar{v}}{1-\bar{u}\bar{v}})\check{R}_{jk}(u)\check{R}_{ij}(v)=
\check{R}_{jk}(\bar{v})\check{R}_{ij}(\bar{u})
\check{R}_{jk}(\frac{\bar{u}-\bar{v}}{1-\bar{u}\bar{v}}),
\ena
introduced and investigated as parameterized Bell matrix \cite{RZWG,ZhLKL}, and will be considered in the next sections as well. The first diagonal check-matrix  is one of the simple solutions to YBE.

\subsection{GYBE  and the integrable models}

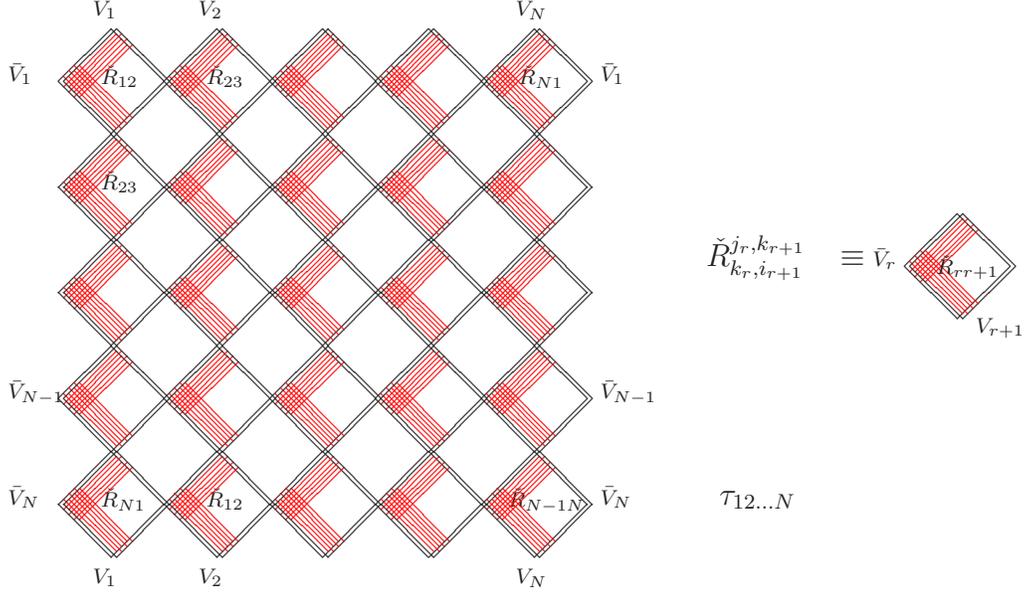
\begin{figure}[t]
\unitlength=10pt
\begin{picture}(100,22)(-1.7,-3)
\newsavebox{\rmatrix}

\sbox{\rmatrix}{\begin{picture}(0,0)(4,4)
\multiput(0,2)(2,-2){2}{\line(1,1){2}}
\multiput(0.2,1.8)(0.2,-0.2){3}{{\color{red}\line(1,1){2}}}
\multiput(2.2,0.2)(0.2,0.2){3}{{\color{red}\line(-1,1){2}}}
\multiput(2,0)(2,2){2}{\line(-1,1){2}}
\end{picture}}

\put(4.6,0.9){\scriptsize{$\check{R}_{N1}$}}
\put(4.6,16.9){\scriptsize{$\check{R}_{12}$}}
\put(8.6,0.9){\scriptsize{$\check{R}_{12}$}}
\put(8.6,16.9){\scriptsize{$\check{R}_{23}$}}
\put(20.4,16.9){\scriptsize{$\check{R}_{N1}$}}
\put(4.6,12.9){\scriptsize{$\check{R}_{23}$}}
\put(20,0.9){\scriptsize{$\check{R}_{N-1N}$}}
\put(4.3,-2){\scriptsize{$V_{1}$}}\put(8.3,-2){\scriptsize{$V_{2}$}}
\put(20.3,-2){\scriptsize{$V_{N}$}}
\put(23.5,1){\scriptsize{$\bar{V}_{N}$}}
\put(23.5,5){\scriptsize{$\bar{V}_{N-1}$}}
\put(23.5,17){\scriptsize{$\bar{V}_{1}$}}
\multiput(7,3)(4,0){5}{\usebox{\rmatrix}}
\multiput(7.2,3)(4,0){5}{\usebox{\rmatrix}}
\multiput(7,19)(4,0){5}{\usebox{\rmatrix}}
\multiput(7.2,19)(4,0){5}{\usebox{\rmatrix}}
\multiput(7,7)(4,0){5}{\usebox{\rmatrix}}
\multiput(7.2,7)(4,0){5}{\usebox{\rmatrix}}
\multiput(7,11)(4,0){5}{\usebox{\rmatrix}}
\multiput(7.2,11)(4,0){5}{\usebox{\rmatrix}}
\multiput(7,15)(4,0){5}{\usebox{\rmatrix}}
\multiput(7.2,15)(4,0){5}{\usebox{\rmatrix}}

\put(1.1,1){\scriptsize{$\bar{V}_{N}$}}
\put(1.1,5){\scriptsize{$\bar{V}_{N-1}$}}
\put(1.1,17){\scriptsize{$\bar{V}_{1}$}}
\put(4.3,19.5){\scriptsize{$V_{1}$}}\put(8.3,19.5){\scriptsize{$V_{2}$}}
\put(20.3,19.5){\scriptsize{$V_{N}$}}
\put(28,1){{$\tau_{12...N}$}}

\multiput(39,12)(0.2,0){2}{\usebox{\rmatrix}}\put(36.2,9.7){\scriptsize{$\check{R}_{rr+1}$}}
\put(27.5,10){$\check{R}_{k_r,i_{r+1}}^{{j}_r,{k}_{r+1}}\;\;\;\equiv$}
\put(33.8,10){\scriptsize{$\bar{V}_{r}$}}
\put(37.7,7.5){\scriptsize{$V_{r+1}$}}
\end{picture}
\caption{Partition function, statistical weights}
\end{figure}

Sometimes there is a vision, that GYBEs in general have no connection with the integrable models.
   When $p\neq 2k$, the $\check{R}$-matrices are inhomogeneous matrices with another structure than there  are used in the case of the usual Algebraic Bethe Ansatz (ABA) \cite{Baxter,KBI}. Let us define the transfer matrix on the cyclic chain with length $N$ and with the vector states $V_{i_k}$ on the $k$-th sites, by means of the $\check{R}$-matrices in GYBE, in the following way
\bea
\tau_{i_1,...i_N}^{j_1...j_N}(u)=\sum_{k_1}...\sum_{k_N}
[\check{R}_{k_{N} i_1}^{j_N k_1}(u)\check{R}_{k_1 i_2 }^{j_1 k_2}(u)....\check{R}_{k_{N-1} i_N}^{j_{N-1} k_{N}}(u)],\label{tt}
\ena
 The matrices $\check{R}$ are defined, as in the Fig. 2 (ii). Note, that  here, in this definition, we do not have auxiliary space in it's usual sense, there are only  quantum spaces which figurate in the formulas.
In the usual definition of the transfer matrix, there is a trace over an auxiliary space \cite{GSA}, and the auxiliary space may differ from the quantum spaces by dimension, meanwhile in (\ref{tt}) the states $\{V_{i_r},\;V_{k_r},\;V_{j_r}\}$, for the given $r$, $(r\in [1,...,N])$, have the same nature. In the Fig.3 the transfer matrices are depicted by the rows. For simplicity, the states $V_{k_r}$, by which the summation is taken in the transfer matrices, are denoted by $\bar{V}_r$, meanwhile the states $V_{i_r,j_r}$ - by $V_r$ (though the vector spaces $V_r$ and $\bar{V}_r$ are similar).
 The $\check{R}$-matrices are presented by the shaded rectangles (rotated by $45$ degrees in comparison to Fig.1,2).

   For simplicity below we shall write the matrices as the functions of one spectral parameter, special cases will be discussed separately.
The transfer matrices commutativity,  $[\tau(u),\tau(v)]=0$, ensures the integrability in ABA.
 Although now the operators $\tau(u)$ and $\tau(v)$  (\ref{tt}) in the product $\tau(u)\tau(v)$ are defined differing by a space shift on the product of the quantum states, the quantum conserved charges, produced by the transfer matrices
  (defined by means of the logarithmic differentials of the $\tau(u)$-operators - $\frac{\partial^n \log[\tau(u)]}{\partial u^n}|_{u=u_0}$),  are the same due to the cyclic  property.    The commutativity of the transfer matrices can be satisfied by  solving of the generalized equations defined on the spaces $V_{i_a}\otimes V_{i_{a+1}}\otimes V_{i_{a+2}}$, $i_{N+1}=i_1$.
\bea \check{R}_{i_a i_{a+1}}(u-v) \check{R}_{i_{a+1} i_{a+2}}(u) \check{R}_{i_a i_{a+1}}(v)=
\check{R}_
{i_{a+1} i_{a+2}}(v)\check{R}_{i_a i_{a+1}}(u)\check{R}_{i_{a+1} i_{a+2}}(u-v). \label{gYBEt}\ena
For the homogeneous case these coincide with the ordinary YBE.

 The transfer matrix we can preset in the graphical form, which we have used in the work \cite{SA}, see the rows in the Fig.3.  The product of the transfer matrices on the periodic $N\times N$ square lattice 
 (having toric periodic conditions):
 $Z_{N\times N}=tr_{i} \prod^{N} \tau$ constitutes  the partition function in  2D statistical physics with the local weights corresponding to the matrices $\check{R}$. A graphical representation of the partition function is depicted on the Fig.3, too. We
apply the checkerboard disposition of the matrices in the square lattice, as in \cite{SA}.

 We shall discuss below  the solutions of the inhomogeneous GYBEs,  concentrating our attention to a concrete specific structure of the $R$-matrices. 

\section{The solutions to high-dimensional GYBE emerged from the usual low-dimensional solutions to YBE}

Let us consider $\check{R}$-matrices, which have non vanishing elements on the diagonal and anti-diagonal positions ($X$-shaped matrices) \cite{RZWG,GHR,Chen}, explored in the context of the quantum information theory as the extensions of $4\times 4$ braiding matrix (the braiding limits of the trigonometric matrix in Eqs. (\ref{r22qt})),
ensuring Bell basis.

 For the general $(N_1)^2\times (N_2)^2$-dimensional case,  there are
 $\mathcal{V}_{\{12\}}$ non-zero elements in such $\check{R}_{N_1 N_2}$-matrices,
\bea
\mathcal{V}_{\{12\}}=\left\{\ba{cc}
2 N_1 N_2&{\mathrm{even}\;N_1\;\mathrm{or}\;N_2}\\
2 N_1 N_2-1&{ \mathrm{odd}\;N_1\;\mathrm{and}\;\mathrm{odd}\;N_2}\ea\right\}
\ena
 i.e. these $\check{R}$-matrices can describe $\mathcal{V}_{\{12\}}$-vertex models in 2d statistical physics.  Considering the spectral parameter dependent GYB equations with such kind matrices will  restrict strongly the area of the solutions.

As it is mentioned, an example of such matrices are the braid group symmetric solutions (see e.g. \cite{GXW})
\bea\label{RM}
\check{R}_M(u)=\frac{1}{2}e^u M^++e^{-u} M^-,\quad M^{\pm}=I\pm M,\quad M^2=-I,\\
M=\left(\ba{ccccc}0&0&\cdots&0&1\\
0&0&\cdots&1&0\\
\vdots&\vdots&\vdots&\vdots&\vdots\\
0&-1&\cdots&0&0\\
-1&0&\cdots&0&0
\ea\right),\;\;
M^{\pm }=\left(\ba{ccccc}1&0&\cdots&0&\pm 1\\
0&1&\cdots&\pm 1&0\\
\vdots&\vdots&\vdots&\vdots&\vdots\\
0&\mp 1&\cdots&1&0\\
\mp 1&0&\cdots&0&1
\ea\right)
\ena
 The matrix $M$ itself is a solution to the rigid (without spectral parameter) GYBE, and $M^{\pm}$-matrices are the braid limits  $lim_{\pm \infty}$ of the mentioned trigonometric solution (\ref{RM}), $\check{R}(u)=\cosh{[u]}I+\sinh{[u]}M$. The more complete trigonometric parametrization of $M^{\pm}$-matrices  we shall discuss in this paper, as extension of (\ref{r22qt}), with an detailed example of $8\times 8$ matrix in Appendix A.1.

  We construct  here the most general GYBE solutions with the matrices having only  diagonal and anti-diagonal non-zero elements, and it occurs, that these are natural extensions of the eight-vertex solutions to ordinary YBE for the high-dimensional matrices. And the mentioned solutions can be  constructed by considering the following ansatz.

 Let us consider the matrix  $\check{R}_{N_1,N_2}$ which acts on the tensor product of the vector spaces $V_{N_1}\otimes V_{N_2}$. The inhomogeneous equations we consider in the following form:
  \bea\label{N123}
  \check{R}_{N_1,N_2}(u,v)\check{R}_{N_2,N_3}(u)\check{R}_{N_1,N_2}(v)
  =\check{R}_{N_2,N_3}(v)\check{R}_{N_1,N_2}(u)\check{R}_{N_2,N_3}(u).
  \ena

  \paragraph{Ordinary GYBE.} When  $N_2=d^{k-p}$ and $N_1=N_3=d^p$, then this set of the equations coincide with $(d,k,p)$-GYBE, providing that the
the  matrices $\check{R}_{{N}_1 {N}_2}$ are
invariant under transposition operation, i.e. the following matrices
$\check{R}_{N_1,N_2}$ and  $\check{R}_{N_2,N_1}$ are equivalent as $(N_1 N_2)\times (N_1 N_2)$ matrices.

  We can enumerate the vector states $|n\rangle$ of the corresponding spaces $V_N$ as follows-  $\{|\mathcal{N}\rangle,|\mathcal{N}-1\rangle ,.|1\rangle,|-1\rangle...,|-\mathcal{N}+1\rangle,|-\mathcal{N}\rangle\}$, for $Dim[V_N]=2\mathcal{N}$, and as $\{|\mathcal{N}\rangle,|\mathcal{N}-1\rangle ,.|0\rangle,...,|-\mathcal{N}+1\rangle,|-\mathcal{N}\rangle\}$, for the odd case: $Dim[V_N]=2\mathcal{N}+1$.
 I.e., here  $\mathcal{N}_i=\left[\frac{Dim[V_{N_i}]}{2}\right]$, $\left[\;\right]$ denotes the integer part of the numbers. The $X$-shaped matrices have the following sub-structures: at $N_1=2\mathcal{N}_1$ and
  $N_1=2\mathcal{N}_1+1$ cases correspondingly: 
\bea
\check{R}_{2\mathcal{N}_1 N_2}=\left(\begin{array}{cc}(\nwarrow)&(\nearrow)\\
(\swarrow)&(\searrow)\end{array}\right),\qquad
\check{R}_{2\mathcal{N}_1+1 N_2}=\left(\begin{array}{ccc}(\nwarrow)&&(\nearrow)\\&(\ast)&\\
(\swarrow)&&(\searrow)\end{array}\right),
\ena
where the slow
 arrows denote diagonal and anti-diagonal $(\mathcal{N}_1)\times(N_2)$ matrices with the following matrix elements
  (disposed from the left to right)
{\begin{small}
\bea
(\nwarrow)=\textsc{Diagonal}\left(\check{R}_{\mathcal{N}_1\;\mathcal{N}_2}^{\mathcal{N}_1\;\mathcal{N}_2},
\check{R}_{\mathcal{N}_1\;\mathcal{N}_2-1}^{\mathcal{N}_1\;\mathcal{N}_2-1},.....
\check{R}_{\mathcal{N}_1\;-\mathcal{N}_2}^{\mathcal{N}_1\;-\mathcal{N}_2},
\check{R}_{\mathcal{N}_1-1\;\mathcal{N}_2}^{\mathcal{N}_1-1\;\mathcal{N}_2},.............
\check{R}_{1\;-\mathcal{N}_2}^{1\; -\mathcal{N}_2}\right),\nn\\
(\nearrow)=\textsc{Anti-Diagonal}\left(\check{R}_{1\;-\mathcal{N}_2}^{-1\; \mathcal{N}_2},\check{R}_{1\;-\mathcal{N}_2+1}^{-1\; \mathcal{N}_2-1},....
\check{R}_{1\;\mathcal{N}_2}^{-1\;-\mathcal{N}_2},\check{R}_{2\;-\mathcal{N}_2}^{-2\;\mathcal{N}_2},............
\check{R}_{\mathcal{N}_1\;\mathcal{N}_2-1}^{-\mathcal{N}_1\;-\mathcal{N}_2+1},
\check{R}_{\mathcal{N}_1\mathcal{N}_2}^{-\mathcal{N}_1\;-\mathcal{N}_2}\right),\nn\\
(\swarrow)=\textsc{Anti-Diagonal}\left(\check{R}_{-\mathcal{N}_1\;-\mathcal{N}_2}^{\mathcal{N}_1\mathcal{N}_2},
\check{R}_{-\mathcal{N}_1\;-\mathcal{N}_2+1}^{\mathcal{N}_1\;\mathcal{N}_2-1},.....
\check{R}_{-\mathcal{N}_1+1\;-\mathcal{N}_2}^{\mathcal{N}_1-1\;\mathcal{N}_2},............
\check{R}_{-1\;\mathcal{N}_2-1}^{1\;-\mathcal{N}_2+1},\check{R}_{-1\;\mathcal{N}_2}^{1\; -\mathcal{N}_2}\right),\nn\\
(\searrow)=\textsc{Diagonal}\left(\check{R}_{-1\;\mathcal{N}_2}^{-1\;\mathcal{N}_2},............
\check{R}_{-\mathcal{N}_1+1\;-\mathcal{N}_2}^{-\mathcal{N}_1+1\;-\mathcal{N}_2},
\check{R}_{-\mathcal{N}_1\;\mathcal{N}_2}^{-\mathcal{N}_1\;\mathcal{N}_2},.....
\check{R}_{-\mathcal{N}_1\;-\mathcal{N}_2+1}^{-\mathcal{N}_1\;-\mathcal{N}_2+1},
\check{R}_{-\mathcal{N}_1\;-\mathcal{N}_2}^{-\mathcal{N}_1\; -\mathcal{N}_2}\right).\nn
\ena
\end{small}}
Here the diagonal and anti-diagonal matrix elements are indexed within the following rules.
   The values of the indexes $\{i,j\}$ of the diagonal $\check{R}_{ij}^{ij}$ elements are decreasing from upper left corner to the lower right position. The first index $i$  runs  from $\mathcal{N}_1$ to $-\mathcal{N}_1$, and for the fixed $i$, the index $j$ goes from $\mathcal{N}_2$ to $-\mathcal{N}_2$.
   The indexes $\{i,j\}$ of the anti-diagonal $\check{R}^{ij}_{-i-j}$ elements are decreasing from the lower left corner to the upper right position. The first index $i$  runs  from $\mathcal{N}_1$ to $-\mathcal{N}_1$, and for fixed $i$ the index $j$ goes from $\mathcal{N}_2$ to $-\mathcal{N}_2$.

  At $N_1=2\mathcal{N}_1+1$, the additional central sub-matrix has the structure $(\ast)_{2\mathcal{N}_2}$ or $(\ast)_{2\mathcal{N}_2+1}$, for the cases $N_2=2\mathcal{N}_2$ and $N_2=2\mathcal{N}_2+1$, respectively:
\bea
(\ast)_{2\mathcal{N}_2}=\!
{
\begin{small}\left(\!\!\begin{array}{ccccc}
\check{R}_{0\;\mathcal{N}_2}^{0\;\mathcal{N}_2}&&&\check{R}_{0\;\;\mathcal{N}_2}^{0\;-\mathcal{N}_2}\\
&{}^{.\;}_{\;.}&{}_{.\;}^{\;.}&\\
&\check{R}_{0\;1}^{0\;1}&\check{R}_{0\;1}^{0\;-1}&\\
&\check{R}_{0\;-1}^{0\;1}&\check{R}_{0\;-1}^{0\;-1}&\\
&{}_{.\;}^{\;.}&{}^{.\;}_{\;.}&\\
\check{R}_{0\;-\mathcal{N}_2}^{0\;\;\mathcal{N}_2}&&&\check{R}_{0\;-\mathcal{N}_2}^{0\;-\mathcal{N}_2}
\end{array}\!\!
\right)\end{small}}, (\ast)_{2N_2+1}=\!{\begin{small}\left(\!\!\begin{array}{ccccc}
\check{R}_{0\;\mathcal{N}_2}^{0\;\mathcal{N}_2}&&&&\check{R}_{0\;\;\mathcal{N}_2}^{0\;-\mathcal{N}_2}\\
&{}^{.\;}_{\;.}&\;&{}_{.\;}^{\;.}&\\
&\;&\check{R}_{0\;0}^{0\;0}&\;&\\
&{}_{.\;}^{\;.}&\;&{}^{.\;}_{\;.}&\\
\check{R}_{0\;-\mathcal{N}_2}^{0\;\;\mathcal{N}_2}&&&&\check{R}_{0\;-\mathcal{N}_2}^{0\;-\mathcal{N}_2}
\end{array}
\!\!\!\right)\end{small}}
\ena

Now let us reformulate the matrices into the superpositions, rewriting the matrix elements in the following way
${[\check{R}_{N_1 N_2}]}_{\pm n \pm p}^{\pm n \pm p}={[\check{r}_{n,p}]}_{\pm \pm}^{\pm \pm}$, $n=\{0/1,...,\mathcal{N}_1\}$, $p=\{0/1,...\mathcal{N}_2\}$,
\bea
\check{R}_{N_1,N_2}=\bigoplus_{n_i=0/1,n_j=0/1}^{\mathcal{N}_1,\mathcal{N}_2} \check{r}_{n_i,n_j}\ena
 with the matrices defined on the minimal spins
$\frac{1}{2}$ and $0$. When $n_i,\;n_j\neq 0$, then the matrix
\bea\check{r}_{n_i,n_j}=\left(\ba{cccc}
{r_{n_i n_j}}_{++}^{++}&0&0&{r_{n_i n_j}}_{++}^{--}\\
0&{r_{n_i n_j}}_{+-}^{+-}&{r_{n_i n_j}}_{+-}^{-+}&0\\
0&{r_{n_i n_j}}_{-+}^{+-}&{r_{n_i n_j}}_{-+}^{-+}&0\\
{r_{n_i n_j}}_{--}^{++}&0&0&{r_{n_i n_j}}_{--}^{--}\ea\right)\ena
constitutes an eight-vertex model defined on the spin-$\frac{1}{2}$ spaces. As the general equations (\ref{N123})
 contain as separate sub-sets the YBEs for each particular matrices $\check{r}_{n_i,n_j}$, hence the corresponding {\bf solutions for the general inhomogeneous eight-vertex models constitute the solutions here}.  And since all the sectorial equations (the parts of the GYBE), which contain different $\check{r}_{n_i,n_j}$-matrices,
\bea \label{YBEIijk}
\check{r}_{n_i,n_j}(u-v)\check{r}_{n_j,n_k}(u)\check{r}_{n_i,n_j}(v)=
\check{r}_{n_j,n_k}(v)\check{r}_{n_i,n_j}(u)\check{r}_{n_j,n_k}(u-v),
\ena
 connect the sub-matrices with different indexes $\{i,j\}$ one with other,  hence all the solutions $\check{r}_{n_i,n_j}(u)$
with any $i,j$ have the same nature (elliptic, trigonometric or rational). And only some variations are possible
 concerning to the constants (model's parameters), as we shall face in the considered examples. The exceptional cases, when there can be disconnected sub-sets of YBEs, are discussed in the Appendix A2, within the Statement III. The possible solutions one can classify as the general XYZ or free-fermionic inhomogeneous matrices which can be formulate by means of the $sl(2)$ operators on the fundamental spin-$\frac{1}{2}$ irreps \cite{Baxter,TF,KS,SHKHAS},  or on the irreps of the $sl_q(2)$ quantum group at general or cyclic cases  \cite{GSA,KS,SHKH113}. The examples, corresponding to the $XXZ$-model or $XX$-model in a transverse field, at ${N}_{1,2}=2,4$  are given in the Appendix A.2.

 Also we can interpret the corresponding spin-chain models with the next-neighboring interactions between the sets of $\frac{1}{2}$-spins of numbers
 $\mathcal{N}_i$ and $\mathcal{N}_{i+1}$, as the models with the interactions between the complex spins at the sites $i,\;i+1$ - $\bigoplus^{\mathcal{N}_{i,i+1}}\frac{1}{2}$. By means of the complex spin operators -  $S_i^a=\sum_{k=1}^{\mathcal{N}_i}(\sigma^a_i)_k$
  and $S_{i+1}^a=\sum_{k=1}^{\mathcal{N}_{i+1}}(\sigma^a_{i+1})_k$ (here $\sigma^a_i$ are the Pauli matrices defined at the $i$-th sites), for the solutions corresponding, as example, to the $XYZ$-model's $\check{r}_{ij}$ matrices, the appropriate 1D spin-model Hamiltonian, $H\approx\frac{d\log[\tau(u)]}{d u}|_{u=0}$,  can be expressed as follows:
  \bea\label{HXYZ}
 H^{\mathcal{N}}_{XYZ}=\sum_{i}H_{i,i+1}=\sum_i  \sum_{k}^{\mathcal{N}_i} \sum_{r}^{\mathcal{N}_{i+1}}\sum_{a=x,y,z}J_a(\sigma^a_i)_k (\sigma^a_{i+1})_r=\sum_i \sum_{a=x,y,z}J_a S_i^a S_{i+1}^a.
  \ena
 %

One can extend the parametrization for the all
  obtained  solutions in this natural way, with arbitrary numbers $\gamma_{i,j}$,
\bea
\check{R}_{{N}_1,{N}_2}(u)=\bigoplus_{p_i=-\mathcal{N}_1,p_j=\mathcal{-N}_2}^{\mathcal{N}_1,\mathcal{N}_2} e^{\gamma_{ij}u}\check{r}_{p_i,p_j}(\alpha_{ij}u).\ena
Of course, there is an ambiguity, connected with the possibility of multiplying the whole $\check{R}$-matrix
 by an overall  function. In the corresponding quantum chain Hamiltonian operators these factors bring us additional
 diagonal summand proportional to $H_0=\sum_k \sum_{i,j}\gamma_{ij}{[I_{p_i}]}_k\otimes {[I_{p_j}]}_{k+1}$. For the even dimensional cases we can formally rewrite it in the following form: $H_0=\sum_k \sum_{i,j}\gamma_{ij}{[(\sigma_z^2)_{p_i}]}_k\otimes {[(\sigma_z^2)_{p_j}]}_{k+1}$.

\paragraph{Elucidations.} One can present other even-dimensional large matrices too, emerged or induced from the well analyzed eight-vertex matrices, satisfying to inhomogeneous GYBEs. For the matrices $\check{R}_{2\mathcal{N}_1\times 2\mathcal{N}_1}$, acting on the product state - $V_{2\mathcal{N}_1}\otimes V_{2\mathcal{N}_2}$, it is natural to propose  matrices $\check{R}_{2\times 2\times \mathcal{N}_1\times \mathcal{N}_1}$ by the following simple formula
\bea
\check{R}_{2\times 2\times \mathcal{N}_1\times \mathcal{N}_2}:V_2\otimes V_2\otimes V_{\mathcal{N}_1}\otimes V_{\mathcal{N}_1}\to V_2\otimes V_2\otimes V_{\mathcal{N}_1}\otimes V_{\mathcal{N}_1},\\\nn
\check{R}_{2\times 2\times \mathcal{N}_1\times N_2}=\check{R}_{2\times 2}\otimes I_{\mathcal{N}_1\times \mathcal{N}_2},\qquad
I_{\mathcal{N}_1\times \mathcal{N}_2}=I_{\mathcal{N}_1}\otimes I_{\mathcal{N}_2}\quad \check{R}_{2\times 2}:V_2\otimes V_2\to V_2\otimes V_2.
\ena
 Then the  following inhomogeneous GYBEs take place, inherited just from the usual YBEs:
\bea
\check{R}_{2\times 2}\check{R}_{2\times 2\times \mathcal{N}_1\times \mathcal{N}_2} \check{R}_{2\times 2}=\check{R}_{2\times 2\times \mathcal{N}_1\times \mathcal{N}_2}\check{R}_{2\times 2}\check{R}_{2\times 2\times \mathcal{N}_1\times \mathcal{N}_2}.\label{22n1n2}
\ena
It is apparent, that another simple definitions also are possible for the matrix $\check{R}_{2\mathcal{N}_1\times 2\mathcal{N}_1}$, induced by one matrix $\check{R}_{2\times 2}$ and the diagonal  matrices, which will act, e.g. on the vectors spaces $V_2\otimes V_{\mathcal{N}_1}\otimes V_2 \otimes V_{\mathcal{N}_2}$ or $V_2\otimes V_{\mathcal{N}_1} \otimes V_{\mathcal{N}_2}\otimes V_2$. The unity matrix can be replaced by diagonal operator with diagonal matrix elements - $\{e^{i\gamma_{1}u}, e^{i\gamma_{2}u},...,e^{i\gamma_{\mathcal{N}}u}\}$.

These operators differ by  dispositions of the matrix elements from those $X$-shaped solutions, for which all the sub-matrices $\check{r}_{ij}$, discussed in this section, coincide one with another, up to overall exponential functions. However by the resulting chain models, they are similar operators (and have the same
eigenvalues), so can be connected by unitary operations. One can check, that the corresponding unitary matrices are not separable/factorisable, and thus promote to more entangling of the states, generated by evolution operators
produced by $X$-shaped $\check{R}$-matrices \cite{LKSL}.

So, we come to the following
\paragraph{Statement I.} The X-shaped even-dimensional matrices, being the solutions of GYBE, are simply the superpositions of the general eight-vertex solutions. The known trigonometric solutions are the particular cases.

And, as an extension regarding to another low-dimensional solutions (e.g. $9\times 9$ chiral Potts models \cite{YPB}-\cite{PAY}, general 15-vertex models \cite{SHKH14}) we can make such
\paragraph{Proposition.} The matrices of dimensions $(p \mathcal{N})^2\times (p \mathcal{K})^2$, constructed by means of $p^2\times p^2$  solutions $\check{r}^{pp}$ to YBE, (and in matrix form reproducing via multiplicative form the shape  of these matrices), via the following representations:
\bea
\check{R}_{p\mathcal{N},p\mathcal{K}}=
\bigoplus_{n_i=1,k_j=1}^{\mathcal{N},\mathcal{K}} \check{r}_{n_i,k_j}^{pp}
\ena
 can be considered as solutions to GYBE. Here instead of the indexes $\{\pm\}$ used in the case of $p=2$, we can use
 the numbers $\{1,..,p\}$. Then, the matrix indexes of the big matrix $[\check{R}_{p\mathcal{N},p\mathcal{K}}]_{ij}^{i'j'}$ are disposed in the following way:
 \bea
 i,i'\in\{{ }_1\mathcal{N},{ }_1(\mathcal{N}-1),...{ }_1 1,{ }_2\mathcal{N},{ }_2(\mathcal{N}-1),...{ }_2 1,....,
 { }_p\mathcal{N},{ }_p(\mathcal{N}-1),...{ }_p 1\},\nn\\
  j,j'\in\{{ }_1\mathcal{K},{ }_1(\mathcal{K}-1),...{ }_1 1,{ }_2\mathcal{K},{ }_2(\mathcal{K}-1),...{ }_2 1,....,
 { }_p\mathcal{K},{ }_p(\mathcal{K}-1),...{ }_p 1\}.\nn
\ena
\subsection{Odd dimensional representations.}
The case, when one of the spaces on which the $\check{R}_{{N}_1 {N}_2}$-matrix acts, has  odd dimension, the situation is changed. 
  Here there are specific equations containing the  indexes $0$.    In comparison to even dimensional
   cases, there is an additional sector in the set of GYBEs, which includes the matrices $\check{r}_{n_i,0}$, $\check{r}_{0,n_j}$ and $\check{r}_{0,0}$ and can be formulated as follows:
 \bea\label{YBEi0k}
\check{r}_{n_i,0}(u-v)\check{r}_{0,n_k}(u)\check{r}_{n_i,0}(v)=
\check{r}_{0,n_k}(v)\check{r}_{n_i,0}(u)\check{r}_{0,n_k}(u-v),\\
\label{YBE0k}
\check{r}_{0,0}(u-v)\check{r}_{0,n_k}(u)\check{r}_{0,0}(v)=
\check{r}_{0,n_k}(v)\check{r}_{0,0}(u)\check{r}_{0,n_k}(u-v),\\\label{YBE0kt}
\check{r}_{0,0}(u-v)\check{r}_{n_k,0}(u)\check{r}_{0,0}(v)=
\check{r}_{n_k,0}(v)\check{r}_{0,0}(u)\check{r}_{n_k,0}(u-v),
\\\label{YBE0}
\check{r}_{0,0}(u-v)\check{r}_{0,0}(u)\check{r}_{0,0}(v)=
\check{r}_{0,0}(v)\check{r}_{0,0}(u)\check{r}_{0,0}(u-v),
\ena
The matrices  $\check{r}_{0,n_i}(u)$ and $\check{r}_{n_i,0}(u)$  have the following $2\times 2$-matrix form
\bea
\check{r}_{0,n_i}=\left(\ba{cc}
{[r_{0n_i}]}_{+}^+&{[r_{0n_i}]}_{+}^-\\
{[r_{0n_i}]}_{-}^+&{[r_{0n_i}]}_{-}^-
\ea\right), \qquad
\check{r}_{n_i,0}=\left(\ba{cc}
{[r_{n_i 0}]}_{+}^+&{[r_{n_i 0}]}_{+}^-\\
{[r_{n_i 0}]}_{-}^+&{[r_{n_i 0}]}_{-}^-
\ea\right).
\ena

The solution of the equation (\ref{YBE0}), taking into account the requirement at the point $u=0$ (coming from the  locality of the models)
\bea
\check{R}_{N_1 N_2}(0)=I_{N_1}\otimes I_{N_2}
\ena
can be easily represented as a function with one arbitrary constant $\alpha$
\bea\label{SYBE0}
\check{r}_{0,0}(u)=e^{\alpha u}.
\ena
 Since  the equations (\ref{YBE0k}, \ref{YBE0kt}) present the following scattering relations on the matrices $[r_{n_i 0}]$ and $[r_{0 n_j}]$ (below we omit the indexes $\{{}_{n_i 0},\; {}_{0n_i}\}$)
\bea \label{rpm}
r_+^+(u)=r_+^+(v)r_+^+(u-v)+r_+^-(v)r_-^+(u-v),\\\nn
r_-^-(u)=r_-^-(v)r_-^-(u-v)+r_-^+(v)r_+^-(u-v),\\\nn
r_+^-(u)=r_+^+(v)r_+^-(u-v)+r_+^-(v)r_-^-(u-v),\\\nn
r_-^+(u)=r_-^+(v)r_+^+(u-v)+r_-^+(v)r_+^+(u-v).
\ena
 Then the solutions are simply  trigonometric  functions
\bea\label{SYBE0k}
r_+^+(u)=r_-^-(u)=\cos[\theta u],\qquad r_+^-(u)=p\sin[\theta u],\quad r_-^+(u)=-\frac{1}{p}\sin[\theta u].
\ena
The parameters $p$ and $\theta$ here are arbitrary numbers (and particularly can be imaginary numbers).

Now it is natural to look for the solutions  ${[\check{r}_{ik}]}_{\pm\pm}^{\pm\pm}$ to the remained equations (\ref{YBEIijk}) as trigonometric ones.   And indeed, the solutions for $[\check{r}_{ij}]$ coincide with the inhomogeneous two-parametric solutions as in (\ref{r22qt}).  Below we prefer to deal with the hyperbolic parameterizations, which has been also used for ordinary YBE  brought in \cite{SHKHAS}:
\bea \label{r-trig}
\check{r}_{ij}(u)=\left(\ba{cccc}
\cosh[\gamma u]&0&0&q \sinh[\beta u]\\
0&\cosh[\beta u]&t \sinh[\gamma u]&0\\
0&{t_1}\sinh[\gamma u]&\cosh[\beta u]&0\\
\frac{t_2}{q}\sinh[\beta u]&0&0&\cosh[\gamma u]\ea\right)
\ena
Here the parameters $\{\gamma,\;\beta,\;q\}$ are independent ones, meanwhile the constants $\{t,\;t_1,\;t_2\}$
are the signs, and the permissible configurations for being the solutions to YBE are the following ones:
\bea
\{t,\;t_1,\;t_2\}=\{\pm 1,\;\pm 1,\;1\},\;\;\{t,\;t_1,\;t_2\}=\{\pm 1,\;\mp 1,\;-1\}.
\ena

So, the results of this sub-section can be formulate in the following
\paragraph{Statement II} The X-shaped odd-dimensional spectral parameter dependent solutions to GYBE  have trigonometric nature.

All these solutions correspond to the 1D quantum Ising model's like spin systems, i.e. free fermionic 1D quantum chains \cite{SHKHAS}, after
 the Jordan-Wigner transformations \cite{JW}. From the other hand, the particular case of the matrix (\ref{r-trig}), when  $\gamma=\epsilon$, is the well observed Bell matrix
 \cite{ZhLKL,RZWG}.

\section{Multi-spectral parametrization for the quantum gate matrices}

 In the quantum information theory the spectral parameter dependent GYBE-solutions realize time evolution $U(t)$ of the quantum states, and it is important to obtain the unitary operator solutions \cite{LKSL}. Note, that the two-parametric trigonometric unitary matrix in (\ref{r22qt}) at $t=1$, $q=e^{i\alpha}$ just
  follows  from the unitary form of the YBE solution (\ref{r-trig}):
 \be \check{R}_U(u)=\frac{\sqrt{2}}{\sqrt{\cosh[2\gamma u]+\cosh[2\beta u]}}\check{R}(u),\qquad t=1,\;t_1=-1,\; t_2=-1,\ee
  after the following transformations of the parameters,
  \bea
  \cos[\theta \tilde{u}]=\frac{\sqrt{2}\cosh[\gamma u]}{\sqrt{\cosh[2\gamma u]+\cosh[2\beta u]}},\qquad
  \sin[\theta \tilde{u}]=\frac{\sqrt{2}\sinh[\beta u]}{\sqrt{\cosh[2\gamma u]+\cosh[2\beta u]}},\\
  \cos[\epsilon \tilde{u}]=\frac{\sqrt{2}\cosh[\beta u]}{\sqrt{\cosh[2\gamma u]+\cosh[2\beta u]}},\qquad
  \sin[\epsilon \tilde{u}]=\frac{\sqrt{2}\sinh[\gamma u]}{\sqrt{\cosh[2\gamma u]+\cosh[2\beta u]}}.\nn
  \ena

   The action of that matrix on the direct product states $|\pm\rangle\otimes |\pm\rangle\equiv|\pm \pm\rangle$, brings to the entangled  states
 \bea\label{RU}
 \check{R}_U\left(\!\ba{c}|+ +\rangle\\|+ -\rangle\\|- +\rangle\\|- -\rangle\ea\!\right)=
 \left(\!\ba{c}\cos[\theta \tilde{u}]\;|+ +\rangle+
 e^{i \alpha}\sin[\theta\tilde{u}]\;|- -\rangle\\\cos[\epsilon \tilde{u}]\;|+ -\rangle+
 \sin[\epsilon\tilde{u}]\;|- +\rangle\\\cos[\epsilon\tilde{u}]\;|-  +\rangle-\sin[\epsilon\tilde{u}]\;|+ -\rangle\\\cos[\theta \tilde{u}]\;|- -\rangle-e^{i \alpha}\sin[\theta \tilde{u}]\;|+ +\rangle\ea\!\right)
 \ena
 In contrast to the known homogeneous case, when $\epsilon=\theta$ (or $\gamma=\beta$), here there are two kind of entangled paired states, with the independent ``concurrences'' - degrees of the entanglement \cite{WKW}, equal to $|\sin[2 \theta \tilde{u}]|$ and $|\sin[2\epsilon \tilde{u}]|$.

The higher generalizations for these matrices  $\check{R}_{N_1 N_2}$, in case, if $N_i=2^{d_i}$, can be considered as complex gates on the tensor products of two-qubit states $\bigotimes^{n} |\pm\rangle$, $n=d_1+d_2$, so called n-qubit states, for obtaining entangled states (such as maximally entangled -  Greenberger-Horne-Zeilinger (GHZ) states or linear cluster states  \cite{CXG,HXW,RBB,ZhG}).

 At $\theta \tilde{u}=\epsilon\tilde{u}=\pm \pi/4,\; \alpha=0$ ($\check{R}=M^{\pm}$) the resulting states constitute the known Bell states - maximal entangled states \cite{BDVSW}. And, as it is noted already, the homogeneous case $\gamma=\beta$; $\epsilon=\theta$, after re-parametrization,  corresponds to the second type solutions (\ref{gYBEss}).

 If the phase $\alpha=0$ and the parameter $\theta$ is time-dependent, then the evaluation of two sub-sets of the vector states $\{|\pm\pm\rangle|$ and $|\pm\mp\rangle\}$ can be described by two independent oscillations with different frequencies, constituting so {\it{Lissajous figures}} in the corresponding parametric spaces.   Meanwhile, there are different visions, how to choose the time-parameter.   The Hamiltonian operator follows just from the Shr\"{o}dinger equation, $H=i\hbar\frac{\partial(U(t))}{\partial t} U^{\dag}$.  Second type YBE solutions (\ref{gYBEss}), $\check{R}(u)=(I+u M)/(\sqrt{1+u^2})$, with time-parameter $u$, give time-dependent Hamiltonian (with time-dependent eigenvalues and time-independent eigenvectors) (e.g. \cite{ZhG}). The trigonometric parametrization \cite{ZhG,ZhLKL} gives time-independent Hamiltonian operator equivalent to operator $M$, with additional deformation parameter $q$ \cite{ZhG,LKSL,J}.  However, one can
 take into account that here (\ref{RU}) there are two independent spectral parameters, and it is possible to define their time dependence by two arbitrary different functions, say as
 \be\epsilon\tilde{u}\approx\frac{\pi}{4}[t],\quad \theta\tilde{u}\approx\frac{\pi}{4}[t-1]. \label{t-ev}\ee
 This choice shows apparently the possibility of independent time-evolutions for two couples of states [$|++\rangle$, $|--\rangle$] and [$|+-\rangle$, $|-+\rangle$]. For the case (\ref{t-ev}) at $t=2n+1$ ($n$ is an integer number) the first states are entirely decoupled, meanwhile the second couple is maximally entangled (for $t=2n$ there is an opposite situation).

  Another Hamiltonian operator is defined by the authors of the work \cite{CXG}, taking the deformation parameter as time-dependent, and the
 usual spectral parameter as time-independent.
 In the case of two-parametric solutions, considered in this paper, if to take time is connected with the parameter $q$, then H is not changed in comparison with the discussions in \cite{CXG}, and the Berry's phases and Berry's sphere for the eigen-states of the Hamiltonian operators are described in the same way.

\section{The graded permutation matrices and multi-parametric generalizations}

 Let us define the following graded permutation matrices: $P_g$ and $P_g^{\tau}$,
 \bea
{[P_g]}_{i_1 i_2}^{j_1 j_2}|i_1\rangle|i_2\rangle=(-1)^{(i_1+j_1)i_1}|j_2\rangle|j_1\rangle,\\
{[P_g^{\tau}]}_{i_1 i_2}^{j_1 j_2}|i_1\rangle|i_2\rangle=(-1)^{(i_1+j_1)i_1}|{j}_2\rangle^{\tau}|{j}_1\rangle^{\tau}
 \ena
Here we use together with the ordinary permutation matrix $P$ (of disposition) of the states, also the colored  permutation ${P}^{\tau}$, realizing the permutation of the states' ``colors'' (types), which for the case $N=2$ means:
\bea
\label{tau-per}
\{1,2\}^{\tau}=\{2,1\}.
\ena
The sign-gradation of the matrix elements could be explained  in a consistent way, introducing the
 graded vector spaces (super-spaces) and graded tensor products (see, as example, the work \cite{GM} and the references therein). Then the basic definitions of the ABA need re-consideration, such as YB equations, $R$-matrices, algebraic relations with monodromy matrices, Lax opeators, transfer matrices. However, YBEs  in the ``check'' formulations, and hence the GYBEs,  remain unchanged. But in this paper we use the ``grading'' term only regarding to the
 ``check'' matrices, without notion of the (anti-)commutation features of the vector states. Formally,
  we exploit below the notion ``parity'' used for the super-spaces, but the fractional numbers in the last sub-section hints that the spaces here may have richer structure (can include para-fermions). The deep understanding of the underlying symmetries we shall leave for further investigation \cite{SHKHoi}, giving here only the recept of the construction of the large matrices.

Note, that at $N=2$ the two-parametric $4\times 4$-matrix (\ref{r-trig}) consists of the sum of the above graded permutation
operators.
\bea
P_g=\left(\ba{cccc}1&&&\\
&&1&\\
&-1&&\\
&&&1\ea\right),\quad P_g^{\tau}=\left(\ba{cccc}&&&1\\
&1&&\\
&&1&\\
-1&&&\ea\right).
\ena
The group relations of these graded colored operators reflect braiding algebra properties
\bea
{[P_g]}_{12} {[P_g]}_{23} [{P_g}]_{12}={[P_g]}_{23}{[P_g]}_{12}{[P_g]}_{23},\\\nn
{[P_g^{\tau}]}_{12}{[P_g^{\tau}]}_{23}{[P_g^{\tau}]}_{12}=
{[P_g^{\tau}]}_{23}{[P_g^{\tau}]}_{12}{[P_g^{\tau}]}_{23},\\\nn
[P_g]_{12}[P_g]_{23}^{\tau}[P_g]_{12}=[P_g]_{23}[P_g^{\tau}]_{12}[P_g]_{23},\\\nn
{[P_g^{\tau}]}_{12}{[P_g]}_{23}{[P_g^{\tau}]}_{12}={[P_g^{\tau}]}_{23}
{[P_g]}_{12}{[P_g^{\tau}]}_{23},\\\nn
{[P_g^{\tau}]}_{12}{[P_g^{\tau}]}_{23}{[P_g]}_{12}=
{[P_g]}_{23}{[P_g^{\tau}]}_{12}{[P_g^{\tau}]}_{23}.
\ena
which means that we have graded and colored representations of the braiding algebra.

Using the described relations let us present the following multi-spectral parametric  solution to the GYBE, for the
homogeneous spaces - also to the usual YBE:
\bea
\check{R}(u;u')=e^{u}P_g+e^{-u}P_g^{\dag}+e^{u'}P_g^{\tau}+e^{-u'}P_g^{\tau\dag},\\
\check{R}_{12}(u-v;u'-v')\check{R}_{23}(u;u')\check{R}_{12}(v;v')=
\check{R}(v;v')_{23}\check{R}(u;u')_{12}\check{R}(u-v;u'-v')_{23}.
\ena
 The operation $\dag$ here denotes the hermitian conjugation. The spectral parameters $\{u,u'\}$ can be presented by means of two constant parameters and one spectral parameter:  $\{u,u'\}$ $\to$ $\{\alpha u,\alpha' u\}$. The generalization of these matrices for the $X$-shaped  matrices $\check{R}_{NK}$, when the large matrices are the combinations  of  $4\times 4$-matrices, is described  in Section 3. As for the case of $4\times 4$ matrix (\ref{r22qt}), the GYB equations $(d,k,p\geq k)$,  with $d^k=N K$, give the general $X$-shaped $\check{R}_{NK}$ trigonometric solution with  independent parameters $\{u,u',...\}$ of number $\mathcal{NK}$.  However the braiding solutions to GYBEs $(d,k,p<k)$ bring to restrictions on the spectral parameters. And the resulting $(N)^2\times(K)^2$ matrices  depend on two spectral parameters (see the examples in the Appendix A.1), and also on additional overall parameters $(\alpha_{ij}u)$:
 \bea
\check{R}_{N,K}(u)=\bigoplus_{p_i,p_j}^{\mathcal{N},\mathcal{K}} e^{\alpha_{ij}u}\check{r}_{p_i,p_j}(u_i;u_j);\quad
\{u_i,u_j\}=\{u,u'\} \quad \mathrm{or} \quad \{u',u\}\ena

 Another generalization,  developed for the homogeneous matrices of $N\times N$ dimensions, we present in the following subsection.

\subsection{The second generalization: colored and graded permutation matrices}

 The basic trigonometric matrix $\check{R}$ (\ref{r-trig}) consists of two YBE solutions $\check{R}^0$ and $\check{R}^{\tau}$, with different spectral parameters:
 \bea
\check{R}(u;u')=\check{R}^{0}(u)+\check{R}^{\tau}(u'). \label{ruux}
\ena
 The graded permutation matrix $P_g$ in
\bea\label{rg}
 \check{R}^{0}(u)=\frac{1}{2}(e^{u}P_g+e^{-u}P_g^{\dag})\!=\!\left(\!\ba{cccc}\cosh{u}&0&0&0\\
0&0&\sinh{u}&0\\0&-\sinh{u}&0&0\\0&0&0&\cosh{u}\ea\!\right),
\ena
as one of the basic components of such decomposition  can be extended to $N\times N$ matrices ${P_{g\;ij}}_{i_1 j_1}^{i_2 j_2}=\delta_{i_1}^{j_2}\delta_{i_2}^{i_2}(-1)^{p_{12}}$, with an undefined  parity $p_{12}$. 

 The next summand $\check{R}^{\tau}$ consists of the colored  graded permutation operators. If the vector space $V$ has the states of the number $N$, i.e. has $N$ colors, then it is possible to construct the colored permutation operators $P_{ij}^{\tau}$ with the corresponding ${ }^{\tau}$ -permutations of the states: $ \{i_1,...i_N\}\to \{i_1^{\tau},...,i_N^{\tau}\}$, $P^{\tau}=P\cdot I^{\tau}$:
\bea
I_{ij}^{\tau}:V_i\otimes V_j\to V_i^{\tau}\otimes V_j^{\tau},\qquad
P_{ij}^{\tau}:V_i\otimes V_j\to V_j^{\tau}\otimes V_i^{\tau}.
\ena
This means, that the permutation operator permutates not only the disposition of the states in the tensor product, but also the colors of the  states. For the simple $N=2$ case the there is one such possibility (\ref{tau-per}):
\bea
\check{R}^{\tau}(u)=\frac{1}{2}(e^{u}P_g^{\tau }+e^{-u}P_g^{\tau \dag})\!=\!\left(\!\ba{cccc}0&0&0&\sinh{u}\\
0&\cosh{u}&0&0\\0&0&\cosh{u}&0\\-\sinh{u}&0&0&0\ea\!\right)\label{rtau}.
\ena
One can verify, that together with the ordinary permutation operators, the general colored $P^{\tau}$-operators, ${P_{ij}^{\tau}}_{i_1 j_1}^{i_2 j_2}=\delta_{i_1}^{j_{\tau 2}}\delta_{i_2}^{i_{\tau 2}}$ also satisfy to the braid group relations.
\bea
P^{\tau}_{ij}P^{\tau}_{jk}P^{\tau}_{ij}=P^{\tau}_{jk}P^{\tau}_{ij}P^{\tau}_{jk}.\label{prr}
\ena
This is the consequence of the fact, that
the colored permutations just are connecting with the representations of the symmetric group $S_N$ (including the subgroup of permutations) and the partition algebra \cite{LKPKur,part}.

Now let us denote the different colored permutations for $N$-dimensional vector spaces as $\tau_i$, $P^{\tau_i}P^{\tau_j}=I^{\tau_{i+j}}$. Note, that the permutation operation of the colors can be either a full permutation, which means that $V^{\tau}_i\neq V_i$ for each state, or a partial permutation for the remaining cases.  Here we discuss only the simple case,
when the colored permutations are identically invertible, i.e.  $(P^{\tau_i})^2=I$, ensuring unitary $\check{R}^{\tau_i}$-operators. The grading operation can be formally introduced in the standard way, by means of the parities: $[P_{ij}^{\tau{g}}]_{i_1 j_1}^{i_2 j_2}=(-1)^{p_{12}}\delta_{i_1}^{j_{2}\tau}\delta_{i_2}^{i_{2}\tau}$.  Here we do not discuss the question of the gradation of the representation states, as well the question of  underlying symmetries (super-symmetries), as it needs separate deep consideration.  We
can see below, that for the parities there are permissible not only integer numbers, but also fractional values, i.e. in general case, the factors $(-1)^{p_{12}}$ just are phases (we are interested mainly with the unitary operators). And for general $N$ also, the matrices constructed
by the Yang-Baxerizations, as in the formulas (\ref{rg}) and (\ref{rtau}), are the solutions to YBE (or GYBE).

Now we can generalize the superposition form of the solution (\ref{ruux}).  The possible trigonometric solutions must be:
 \bea
 \check{R}^{a}(u)=\frac{1}{2}(e^{\alpha_a u}P_g^{\tau_a}+e^{-\alpha_a u}P_g^{\tau_a\dag}),\\
  \check{R}^{ab}(u)=\frac{1}{2}(e^{\alpha_a u}P_g^{\tau_a}+e^{-\alpha_a u}P_g^{\tau_a \dag})+\frac{1}{2}(e^{\alpha_b u}P_g^{\tau_b}+e^{-\alpha_b u}P_g^{\tau_b \dag})\equiv \check{R}_{a}(u)+\check{R}_{b}(u),\\
  ...\nn
  \\
  \check{R}^{a...c}=\check{R}^{a}(u)+\cdots +\check{R}^{c}(u).
 \ena
The constraints followed from the GYBE give the relations only on the grading rules and also  on the parameters $\alpha_a, \alpha_b,...$, which in some cases coincide one with another. There can be also situation, that for the given combination of the colored matrices $\{a,...,a'\}$ the corresponding matrix $\check{R}^{a...a'}$ will not have appropriate set of the parameters and parities satisfying to YB or GYBE equations. We can extend our research for the cases, when the colored permutations satisfy to $P^{\tau_1}...P^{\tau_k}=I$. Then in order to check that the  unitary operators, constructed by linear superpositions of these operators, satisfy the braid relations (or GYBE), it must be analyzed the appropriate {\it sequence of the colored multiple braiding relations}.

At $N=2$ the solution can contain two arbitrary spectral parameters, i.e two independent parameters $\alpha_{a,b}$.
For the general case of $N$ the situation is much more complicated, there are different possibilities of the gradings,
  but less rich possibilities for the multi-parametric parameterizations.

 The examples of the solutions with the full permutation operators
  for   $4^2\times 4^2$-matrix one can find in the next Subsections. We shall see, that the permissible parities are different for different kind of GYBEs.

\subsection{$16\times 16$-matrices, solutions to YBE constructed with graded and colored permutation operators}

In this subsection we consider of $16\times 16$-matrices, which can be considered
 as the solutions to  GYBE $(2,4,2)$, (which is just the ordinary YBE $(4,2,1)$, as we are considering the homogeneous matrices), or GYBE $(2,4,1)$, as well GYBE $(2,4,3)$.

 In this particular case the   whole set of the colored g-permutations consists of the following four full permutation operators $\left(P^e\right)^2=1$, acting on the four-dimensional space with the vectors  states $|v_i\rangle$, $i=1,2,3,4$, and equipped with corresponding parities $p_e$:
\bea
{[{P^0_g}]}_{i_1 i_2}^{j_1 j_2}
&=&(-1)^{p_0}\delta_{i_1}^{j_2}\delta_{i_2}^{j_1}, \qquad \{|v_i\rangle\}^0=\{v_1,v_2,v_3,v_4\},\\
 {[{P^a_g}]}_{i_1 i_2}^{j_1 j_2}
 &=&(-1)^{p_a}\delta_{i_1}^{j_2a}\delta_{i_2}^{j_1a}, \qquad
 \{|v_i\rangle\}^a=\{v_2,v_1,v_4,v_3\}, \\
 {[{P^b_g}]}_{i_1 i_2}^{j_1 j_2}
 &=&(-1)^{p_b}\delta_{i_1}^{j_2b}\delta_{i_2}^{j_1b}, \qquad  \{|v_i\rangle\}^b=\{v_3,v_4,v_1,v_2\}, \\
 {[{P^c_g}]}_{i_1 i_2}^{j_1 j_2}
 &=&(-1)^{p_c}\delta_{i_1}^{j_2c}\delta_{i_2}^{j_1c}, \qquad  \{|v_i\rangle\}^c=\{v_4,v_3,v_2,v_1\}.
\ena
  For the homogeneous case the parity matrices $(-1)^{p_e}$,  $e=0,a,b,c$, are just the signs, and here they
  are chosen in a rather general form, only keeping the requirement that the diagonal matrix elements must be $1$,
 \bea
 {[p_e]}_{i_1 i_2}^{j_1 j_2}=\{0,1\},\qquad [p]_{i_1 i_2}^{i_1 i_2}=0.
 \ena
 %
  %
  The matrices $P^{e}_g$ satisfy to constant YBE  (or GYBE). The simplest spectral dependent matrices with the constructed permutation matrices  are the following ones
 \be
 \check{R}^e(u)=\frac{1}{2}\left(e^{u}P^e_g+e^{-u}P^{e\dag}_g\right),
  \ee
  where the sign parameters are not fixed (they are arbitrary ones). The next matrices are $\check{R}^{e'e''}(u,v)=R^{e'}(u)+R^{e''}(v)$. For such double cases $\{e',e''\}$ the GYBE put definite constraints on the  parameters $[{p_e}]_{i_1 i_2}^{j_1 j_2}$, (in most cases the second  indexes $\{i_2,j_2\}$ in $[{p_e}]_{i_1 i_2}^{j_1 j_2}$ for the given permutation matrices can be omitted).

   Let us construct  linear superposition of the  operators  $\check{R}^e(u)$, applicable in the integrable models:
  \bea
\check{R}^{4\times 4}(\alpha_0,\alpha_a,\alpha_b,\alpha_c;u)= \check{R}^0(\alpha_0 u)+\check{R}^a(\alpha_a u)+
\check{R}^b(\alpha_b u)+\check{R}^c(\alpha_c u), \quad
\check{R}^{4\times 4}(0)=I.
 \ena
The checking of the YBEs, which turn into the  constraints on
 the parities  $p_e$ for such linear combination of the permutation matrices, shows that there are many possibilities for the solutions of $p_e$-s.
 The choice of the signs or phases are different in different configurations of the GYBEs. As example, the elements of the $p_e$-matrices are different for the solutions to $(2,4,2)$ and $(2,4,1)$.
  One set of the solutions, which we consider in this section, has rather symmetric constraints on $p_e$, which look like as:
 \bea
 {[p_e]}_{i_1 i_2}^{j_1 j_2}\equiv {[p_e]}^{j_1}_{i_1},\quad {[p_e]}_{j_1}^{i_1}={[p_e]}_{i_1}^{j_1}+1,\quad i_1\neq j_1,\nn\\
 {[p_e]}_{j_1}^{i_1}={[p_{e'}]}_{j_1}^{i_1},\quad \forall \{e,e'\},\label{signs}\\
 {[p_e]}_{4}^{1}={[p_e]}_{3}^{2}+1,\quad {[p_e]}_{2}^{1}={[p_e]}_{4}^{3}+1,\quad
 {[p_e]}_{3}^{1}={[p_e]}_{4}^{2}+1.\nn
 \ena
Let us represent the operator  $R^{4\times 4}(\alpha_0,\alpha_a,\alpha_b,\alpha_c;u)$ with a proper sign configuration, satisfying to (\ref{signs}), in general $16\times 16$  matrix form, with arbitrary parameters $\alpha_e$,

 {\begin{small}
 \bea
& f_u^e=\sinh{[\alpha_e u]},\quad g_u^e=\cosh{[\alpha_e u]},\qquad \check{R}^{4\times 4}(\alpha_0,\alpha_a,\alpha_b,\alpha_c;u)=&\label{rfg}\\
 &\nn&\\\nn
 &\left(\ba{cccccccccccccccc}
 g_u^0&&&&&f_u^a&&&&&f_u^b&&&&&\!-\!f_u^c\\
 &g_u^a&&&f_u^0&&&&&&&f_u^c&&&\!-\!f_u^b&\\
 &&g_u^b&&&&&f_u^c&f_u^0&&&&&\!-\!f_u^a&&\\
 &&&g_u^c&&&f_u^b&&&f_u^a&&&\!-\!f_u^0&&&\\
 &\!-\!f_u^0&&&g_u^a&&&&&&&f_u^b&&&f_u^c&\\
 \!-\!f_u^a&&&&&g_u^0&&&&&f_u^c&&&&&f_u^b\\
 &&&\!-\!f_u^b&&&g_u^c&&&f_u^0&&&f_u^a&&&\\
 &&\!-\!f_u^c&&&&&g_u^b&f_u^a&&&&&f_u^0&&\\
 &&\!-\!f_u^0&&&&&\!-\!f_u^a&g_u^b&&&&&\!-\!f_u^c&&\\
&&&\!-\!f_u^a&&&\!-\!f_u^0&&&g_u^c&&&\!-\!f_u^b&&&\\
\!-\!f_u^b&&&&&\!-\!f_u^c&&&&&g_u^0&&&&&\!-\!f_u^a\\
 &\!-\!f_u^c&&&\!-\!f_u^b&&&&&&&g_u^a&&&\!-\!f_u^0&\\
 &&&f_u^0&&&\!-\!f_u^a&&&f_u^b&&&g_u^c&&&\\
 &&f_u^a&&&&&\!-\!f_u^0&f_u^c&&&&&g_u^b&&\\
 &f_u^b&&&\!-\!f_u^c&&&&&&&f_u^0&&&g_u^a&\\
 f_u^c&&&&&\!-\!f_u^b&&&&&f_u^a&&&&&g_u^0 \ea
 \right)&
 \ena
 \end{small}}
These matrices, taking into account the normalizing factor $\left(\!{\sqrt{1\!+\!(\!f_u^0)^2+\!(\!f_u^a)^2+\!(\!f_u^b)^2+\!(\!f_u^c)^2}}\right)^{-1}$, are unitary for independent values of the parameters $\alpha_e$, but however the GYBEs (YBEs) put the restrictions on their values.
 Then one can directly check that the following operators are solutions to YBE.
  {\begin{small}
 \bea\label{ralpha0}
\check{ R}(\alpha,\alpha,\alpha,\alpha;u),\quad
 \check{R}(\alpha,0,\alpha, 0;u),\quad \check{R}(0,\alpha,0,\alpha;u),\quad \check{R}(\alpha,0,0,\alpha;u),\nn\\
\check{R}(0,\alpha,\alpha, 0;u),\quad \check{R}(\alpha,\alpha,0, 0;u),\quad  \check{R}(0,0,\alpha,\alpha;u)
 \ena
  \end{small}}

We differentiate two type of the solutions. The first type, considered above, admits the property $\check{R}(0)=I$. For these solutions, the integrable local models can be constructed by standard way (see the next subsection).

\paragraph{Exceptional solutions $\sum_e\rho_e\check{R}^e(u),\quad \prod_e \rho_e=0.$} The next possible $16\times 16$ solutions can be considered by partial sums of the matrices $\check{R}^e(u)$, $e=\{0,a,b,c\}$. By direct derivation it is easy to verify, that each matrix $\check{R}^e(u)$ is a solution to YBE. Moreover,  matrices with the non-vanishing elements,  disposed so, as in the matrices $\check{R}^e(u)$, are the solutions to YBE with any constants or functions. That is, the generalized matrices
 $$\check{R}_G^e(u)=[r^e]_{i_1 i_2}(u)\delta_{i_1}^{j_2e}\delta_{i_2}^{j_1e},$$
  with any functions $[r^e(u)]_{i_1 i_2}=[r^e(u)]_{i_1e,j_2e}$, are the solutions to YBE. But already any combination of two operators $[\check{R}^e(u)+\check{R}^{e'}(u')]$ brings to definite functions and signs. Thus, any matrix (\ref{rfg}), with two sets of
vanishing elements $\{f_{e''}=0,\;g_{e''}=0\}$, $\{f_{e'''}=0,\;g_{e'''}=0\}$, $e'',e'''\neq e,e'$,  satisfies to YBE, with the following possible values of the
parameters $\{\alpha_e,\alpha_{e'}\}=\{\alpha,\alpha\}$, $\{\alpha_e,\alpha_{e'}\}=\{\alpha,0\}$ or  $\{\alpha_e,\alpha_{e'}\}=\{0,\alpha\}$.

Let us present some exact solutions and the corresponding four-dimensional spin-spin Hamlitonian operators.

 \subsubsection{$\alpha_e=\alpha$}

For the homogeneous solutions,  all parameters $\alpha_e$ equal one to another: $\alpha_e=\alpha$. We fix $\alpha=1$.
We can represent this large matrix in the following form, by means of  $4\times 4$-operators: by $Z_4$-operators $\bar{M}_{x,y,z}$, $(\breve{M}^4=1)$, and by $Z_2$-operators $\bar{M}_{x,y,z}$, $(\breve{M}^2=1)$ (below $I_n$ is the $n\times n$ unity operator, and the signs $\varepsilon_{x,y,z}^2=1$ reproduce the solutions for the parity matrices  permissible by the relations (\ref{signs}):
{\begin{small}
\bea\label{r44}
\check{R}_{44}(u)=\cosh{[u]}I_4\otimes I_4+\sinh{[u]}\left(\varepsilon_x\breve{M}_x\otimes\bar{M}_x+
\varepsilon_y\breve{M}_y\otimes\bar{M}_y+
\varepsilon_z\breve{M}_z\otimes\bar{M}_z\right),
\ena
\bea
\check{R}_{44}(u)=\left(\ba{cccc}
\cosh{[u]}I_4&\varepsilon_x\sinh{[u]}\bar{M}_x&\varepsilon_y\sinh{[u]}\bar{M}_y&-\varepsilon_z\sinh{[u]}\bar{M}_z\\
-\varepsilon_x\sinh{[u]}\bar{M}_x&\cosh{[u]}I_4&\varepsilon_z\sinh{[u]}\bar{M}_z&\varepsilon_y\sinh{[u]}\bar{M}_y\\
-\varepsilon_y\sinh{[u]}\bar{M}_y&-\varepsilon_z\sinh{[u]}\bar{M}_z&\cosh{[u]}I_4&-\varepsilon_x\sinh{[u]}\bar{M}_x\\
\varepsilon_z\sinh{[u]}\bar{M}_z&-\varepsilon_y\sinh{[u]}\bar{M}_y&\varepsilon_x\sinh{[u]}\bar{M}_x&\cosh{[u]}I_4\ea\right)
\ena
\bea
&\bar{M}_x=\left(\ba{cccc}0&1&0&0\\
1&0&0&0\\
0&0&0&1\\
0&0&1&0\ea\right),\;\; \bar{M}_y=\left(\ba{cccc}0&0&1&0\\
0&0&0&1\\
1&0&0&0\\
0&1&0&0\ea\right),\;\;
\bar{M}_z=\left(\ba{cccc}0&0&0&1\\
0&0&1&0\\
0&1&0&0\\
1&0&0&0\ea\right),&\\&
\breve{M}_x=\left(\ba{cccc}0&1&0&0\\
-1&0&0&0\\
0&0&0&-1\\
0&0&1&0\ea\right),\;\; \breve{M}_y=\left(\ba{cccc}0&0&1&0\\
0&0&0&1\\
-1&0&0&0\\
0&-1&0&0\ea\right),\;\;\quad \breve{M}_z=\left(\ba{cccc}0&0&0&-1\\
0&0&1&0\\
0&-1&0&0\\
1&0&0&0\ea\right)&
\ena
\end{small}}
The matrix (\ref{rfg}) corresponds to the case with the signs $\varepsilon_{x,y,z}=1$. We see, that the all matrices $\breve{M}_i$ have only two eigenvalues, so the  matrices are
constructed rather by the generators of $Z_2\otimes Z_2$ (or $sl_2 \times sl_2$) and not $Z_4$. The matrices $\breve{M},\; \breve{M}$ can be presented by means of the tensor product of the Pauli matrices $\sigma_x=\left({}^0_1{}^1_0\right)$, $\sigma_y=\left({}^0_{i}{}^{-i}_0\right)$, $\sigma_z=\left({}^1_0{}^0_{-1}\right)$:
\bea
\bar{M}_x=I_2\otimes \sigma_x,\quad \bar{M}_y=\sigma_x\otimes I_2,\quad \bar{M}_z=\sigma_x\otimes \sigma_x,\\
\breve{M}_x=i \sigma_z\otimes \sigma_y,\quad\breve{M}_y=i\sigma_y\otimes
  I_2,\quad\breve{M}_z=-i\sigma_x\otimes \sigma_y.
\ena
We see, that $\bar{M}$-operators are commutative, and the operators $i \breve{M}$ reproduce the algebra
 of $sl_2$.
 Then the
 corresponding quantum spin-chain Hamiltonian operator for the matrix $\check{R}_{44}(u)$ reads as
 \bea
 H_{44}\approx J
 \sum_k\left([\sigma_z]_k\otimes [\sigma_y]_k\otimes [I_2]_{k+1}\otimes [\sigma_x]_{k+1}+\right.\\\nn
 [\sigma_y]_k\otimes [I_2]_k\otimes
 [\sigma_x]_{k+1}\otimes [I_2]_{k+1}-
 \left.[\sigma_x]_k\otimes [\sigma_y]_k\otimes
 [\sigma_x]_{k+1}\otimes [\sigma_x]_{k+1}
 \right).
 \ena
The model is invariant under $Z_2\otimes Z_2$ symmetry: there is a conserved charge - $\prod_k [\sigma_z\otimes \sigma_z]_k$.  

\subsubsection{$\alpha_0=\alpha_a\equiv 1,\quad \alpha_e=0,\; e=b,c$}

Let us represent the expression by means of the spin operators for  one of the  next unitary matrices (\ref{ralpha0}, \ref{rfg}), with non vanishing parameters $\alpha_0=\alpha_a=1$ ($\alpha_b=\alpha_c=0$).
\bea\nn
\check{R}_{44}(1,1,0,0;u)=
\frac{(\cosh[u]+1)}{2}I_2\otimes I_2\otimes I_2\otimes I_2+
\frac{(\cosh[u]-1)}{2}\sigma_z\otimes I_2\otimes \sigma_z
\otimes I_2+\\
\frac{i\sinh[u]}{2} \left(\sigma_y\otimes I_2\otimes \sigma_x
\otimes I_2-\sigma_x\otimes I_2\otimes \sigma_y
\otimes I_2+\sigma_z\otimes \sigma_y\otimes I_2\otimes \sigma_x
+I_2\otimes \sigma_y\otimes \sigma_z\otimes \sigma_x \right).
\ena

\subsection{$16\times 16$ matrices as the solutions to GYBE $(2,4,1)$}

For these cases the parities may have fractional values, i.e.
 \bea
 {[p_e]}_{i_1 i_2}^{j_1 j_2}=\{0,1/2,1\},\qquad [p]_{i_1 i_2}^{i_1 i_2}=0.
 \ena

  One of the trigonometric solutions to GYBE $(2,4,1)$ can be taken in the matrix form (\ref{r44}) (unitary properties need careful consideration), with the constraints as in Eqs. (\ref{signs}), only the phase factors now have the property:
 \be\varepsilon_x^2=1,\quad \varepsilon_y^2=-1,\quad \varepsilon_z^2=-1.\ee
 The corresponding operators are
\bea\label{vv}
\check{R}_{2222}(u)=\cosh{[u]}I_2\otimes I_2\otimes I_2\otimes I_2+\sinh{[u]}\left(\pm i \sigma_z\otimes\sigma_y\otimes I_2\otimes\sigma_x+\right.\\\nn
\left.
\pm \sigma_y\otimes I_2\otimes\sigma_x\otimes I_2
\pm \sigma_x\otimes\sigma_y\otimes \sigma_x\otimes\sigma_x\right).
\ena

In the inhomogeneous GYBE formulation (\ref{YBEII}) these matrices act as $\check{R}_{2^3 2}(u)$ and $\check{R}_{2 2^3 }(u)$.
\section{Summary}

We have investigated the generalized Yang-Baxter equations, developed for deriving the unitary quantum gates. 
 The main goal of this paper is the  construction of new solutions to GYBE, which could be interesting also in the theory of the integrable models.

  Two kind of the multi-state operators  of arbitrary dimensions are considered. The first version (X-shaped) of the solutions, for even dimensional $(2\mathcal{N}_1)^2\times (2 \mathcal{N}_2)^2$ matrices, just turns into the
  consideration of $\mathcal{N}_1\times \mathcal{N}_2$ YBEs with eight-vertex type solutions. The odd dimensional cases for the first version admit only trigonometric spectral parametric solutions.
  The obtained large matrices correspond to the mixed interactions between the sets of the $\frac{1}{2}$-spins (only including even-dimensional states) and $0$-spins (including also odd-dimensional
 cases) situated on the sites of the cyclic chains. Here, in contrast to the integrable models on the composite representations of the quantum algebras, discussed in \cite{SHKKH09,ShKh18}, the Hamiltonian operators are completely separable on direct sums of independent identical interactions. This series is discussed in Section 3 and Appendix (with even dimensional examples), the odd dimensional case (as well more non-homogeneous solutions to GYBE) in details  will be separately investigated in \cite{SHKHoi}.

  The second versions of the solutions, which are constructed by means of the graded
   and colored permutation operators, also admit trigonometric Yang-Baxterization, and here the GYBEs become the equations on the parity-matrices. The possibility of the choosing of the parities, we see, that is not unique. The second kind extensions  can also be formulated in terms of the
 spin  interactions. Apparently, such Hamiltonian operators do not possess ordinary $sl(2)$-symmetry, but however can have  symmetries of the algebra's deformations.   Such questions we are addressing to the next works. In the same time the topological properties of the underlying integrable models could be investigated, especially the obtained fractional parities may be related to the anyon  (parafermion) statistic, considering in topological quantum computation \cite{FKL,Kit,z3}.   Also the connection with the higher dimensional integrable models needs to be considered \cite{SHKAS-IP,SHKHgen}.

   And another novelty of our research is the multi-spectral parameter Yang-Baxterization of unitary X-shaped series of the rigid GYBE solutions $M^{\pm}$ (extensions of the known quantum Bell type matrices \cite{ZhG}), examples are presented in Appendix A.1. For the usual two-qubit gates the evolution of the states by such unitary transformations is briefly discussed in Section 4, which brings to two kind of entanglement's degrees, relying on two independent spectral-parameters. Note, that the spectral-parameter dependent  Bell matrices just are the multi-parametric solutions of separable GYBEs $(d,k,p)$, $p\geq k$ (\ref{r22qt}), the equations with $k>p$ put restrictions on the possible sets of the independent parameters, see Appendix A.1.

 \paragraph{Acknowledgements.} The work was supported by the Science Committee of RA, in the frames of research projects  20TTAT-QTa009, 20TTWS-1C035 and 21AG-1C024.

\setcounter{section}{0}
\renewcommand{\thesection}{\Alph{section}}\Alph{section}
\renewcommand{\theequation}{\Alph{section}.\arabic{equation}}
\addtocounter{section}{0}\setcounter{equation}{0}

\section{Solutions to homogeneous and inhomogeneous GYBE with $2^3\times 2^3$-matrices: $(2,3,1)$, $(\{2\},\{3,2\},\{1,2\})$.}

 Here we are considering $X$-shaped spectral parameter dependent solutions  defined on the space $V_2\otimes V_2\otimes V_2$ in the usual GYBE formulation $(2,3,1)$ and $(2,3,2)$. In the inhomogeneous ii-type  GYBE formulation these are the equations $(\{2,4,2\},\{2\},\{1\})$ or $(\{4,2,4\},\{2\},\{2\})$ with the matrices $\check{R}_{24}$ and $\check{R}_{42}$ acting on the spaces $V_2\otimes V_4$ or $V_4\otimes V_2$. Such matrices satisfy to the following inhomogeneous equations:
 \bea
 \check{R}_{24}(u-v) \check{R}_{42}(u)  \check{R}_{24}(v) = \check{R}_{42}(v)  \check{R}_{24}(u)
  \check{R}_{42}(u-v),\nn\\
   \check{R}_{42}(u-v) \check{R}_{24}(u)  \check{R}_{42}(v) = \check{R}_{24}(v)  \check{R}_{42}(u)
  \check{R}_{24}(u-v).
\ena
 These matrices also can be considered together with the matrices defined on $V_2\otimes V_2$ in the  inhomogeneous GYBEs of  ii-type $(\{2,2,4\},\{2\},\{1\})$ and $(\{4,2,2\},\{2\},\{1\})$
\bea
\check{R}_{22}(u-v)  \check{R}_{24}(u)  \check{R}_{22}(v) =
  \check{R}_{24}(u-v)  \check{R}_{22}(u-v)  \check{R}_{24}(u-v),\nn\\
  \check{R}_{42}(u-v)  \check{R}_{22}(u)  \check{R}_{42}(v) =
  \check{R}_{22}(u-v)  \check{R}_{42}(u-v)  \check{R}_{22}(u-v).\label{424}
 \ena
These equations also can be formulated as the type-i inhomogeneous GYBEs $(\{2\},\{2,3\},\{2,1\})$ or $(\{2\},\{3,2\},\{1,2\})$

If  $4\times 4$-matrix constitutes one of the known eight-vertex model's solutions, these $8\times 8$-matrices consist of two eight-vertex type solutions with the same   structure.
 Below we consider explicitly two type solutions, for which there are unitary parameterizations (one-parametric $4\times 4$-matrices are detailed considered in \cite{ZhLKL}), so these matrices can be used both in the theory of integrable models, and in the quantum information theory.

\subsection{ Two-parametric trigonometric matrices}

 As example, for the trigonometric two-parametric matrix $\check{R}_{22}(\alpha_0 u,\alpha_x u)$, the possible solutions are the following ones - $\check{R}_{42}(u)$ (\ref{r42}) and $\check{R}_{24}(u)$ (\ref{r24}), where
 for simplicity we defined $\{u_i=\alpha_i u\}$, $i=1,2,3,4$:
{\begin{small}
\bea\label{r42}
\left(\!\!\begin{array}{cccccccc}\cosh{u_1}&&&&&&&\sinh{u_2}\\
&\cosh{u_2}&&&&&\sinh{u_1}&\\
&&e^{\alpha u}\cosh{u_3}&&&e^{\alpha u}\sinh{u_4}&&\\
&&&e^{\alpha u}\cosh{u_4}&e^{\alpha u}\sinh{u_3}&&&\\
&&&-e^{\alpha u}\sinh{u_3}&e^{\alpha u}\cosh{u_4}&&&\\
&&-e^{\alpha u}\sinh{u_4}&&&e^{\alpha u}\cosh{u_3}&&\\
&-\sinh{u_1}&&&&&\cosh{u_2}&\\
-\sinh{u_2}&&&&&&&\cosh{u_1}
\end{array}
\!\!\right)
\ena
\bea\label{r24}
\left(\!\!\begin{array}{cccccccc}\cosh{u_1}&&&&&&&\sinh{u_2}\\
&e^{\alpha u}\cosh{u_3}&&&&&e^{\alpha u}\sinh{u_4}&\\
&&e^{\alpha u}\cosh{u_4}&&&e^{\alpha u}\sinh{u_3}&&\\
&&&\cosh{u_2}&\sinh{u_1}&&&\\
&&&-\sinh{u_1}&\cosh{u_2}&&&\\
&&-e^{\alpha u}\sinh{u_3}&&&e^{\alpha u}\cosh{u_4}&&\\
&-e^{\alpha u}\sinh{u_4}&&&&&e^{\alpha u}\cosh{u_3}&\\
-\sinh{u_2}&&&&&&&\cosh{u_1}
\end{array}
\!\!\right)
\ena\end{small}}
The permissible values of the solutions' parameters  are the followings:
\bea
\{\alpha_1,\alpha_2,\alpha_3,\alpha_4\}=\{\alpha_0,\alpha_x,\alpha_0,\alpha_x\},\quad
\{\alpha_1,\alpha_2,\alpha_3,\alpha_4\}=\{\alpha_0,\alpha_x,\alpha_x,\alpha_0\},\\
\{\alpha_1,\alpha_2,\alpha_3,\alpha_4\}=\{\alpha_x,\alpha_0,\alpha_0,\alpha_x\},\quad
\{\alpha_1,\alpha_2,\alpha_3,\alpha_4\}=\{\alpha_x,\alpha_0,\alpha_x,\alpha_0\}.
\ena

The possible variations of the positions of the parameters come from the fact, that in general for the two-parametric $4\times 4$ matrix 
 the following non-homogeneous YBE take place:
\bea \label{ra0a}
\check{R}(\alpha_0,\alpha_x;u-v)\check{R}(\alpha_x,\alpha_0; u)
\check{R}(\alpha_0,\alpha_x; v)=\check{R}(\alpha_x,\alpha_0; v) \check{R}(\alpha_0,\alpha_x; u)
\check{R}(\alpha_x,\alpha_0; u-v)
\ena
together with usual homogeneous equations
\bea
\check{R}(\alpha_0,\alpha_x; u-v)\check{R}(\alpha_0,\alpha_x; u)\check{R}(\alpha_0,\alpha_x; v)=
\check{R}(\alpha_0,\alpha_x; v)\check{R}(\alpha_0,\alpha_x; u)\check{R}(\alpha_0,\alpha_x; u-v).
\ena
As the constants  of YBE solutions like $q$ in (\ref{r22qt}) play definite role as in the theory of integrable models, also in the quantum information theory (and define the deformations of the braiding operators) \cite{J,ChXG}, let us present the richer version of the solutions to (\ref{ra0a}), with different independent deformation constants
$(q,q',t)$:
\bea
\check{R}(\alpha_0,\alpha_x,q,t;u)=\left(\!\ba{cccc}
\cosh[\alpha_0 u]&0&0&q\sinh[\alpha_x u]\\
0&\cosh[\alpha_x u]&t \sinh[ \alpha_0 u]&0\\
0&\frac{-\sinh[\alpha_0 u]}{t}&\cosh[ \alpha_x u]&0\\
\frac{-\sinh[\alpha_x u]}{q}&0&0&\cosh[\alpha_0 u]\!\ea\right)
\ena
\bea \label{ra0at}
\check{R}(\alpha_0,\alpha_x,q,t;u-v)\check{R}(\alpha_x,\alpha_0,q',\frac{q}{q't}; u)
\check{R}(\alpha_0,\alpha_x,q,t; v)=\\\nn
\check{R}(\alpha_x,\alpha_0,q',\frac{q}{q't}; v) \check{R}(\alpha_0,\alpha_x,q,t; u)
\check{R}(\alpha_x,\alpha_0,q',\frac{q}{q't}; u-v)
\ena

In the solutions (\ref{r24}, \ref{r42}) we have omitted the constants like  $q,t$, (deformation parameters of braid group) but in general case they also exist.

\subsection{$XXZ$-type $8\times 8$ matrices}

As another example let us present the matrices $\check{R}_{24}(u)^{\pm}$ and $\check{R}_{42}(u)^{\pm}$ induced from the homogeneous $XXZ$-model's matrix ($\check{R}_{22}^{+}(u)$) and the matrix $\check{R}_{22}^{-}(u)$ describing the $XX$-model at the
transverse field \cite{Baxter,ZhLKL,SHKHAS}
{\begin{small}\bea\label{r22pm}&\check{R}_{22}^{\pm}(u,u_0,\gamma)
=\left(\begin{array}{cccc}\frac{\sinh{[u\!+\!u_0]}}{\sinh{[u_0]}}&&&\\
&1&\frac{e^{\gamma}\sinh{[u]}}{\sinh{[u_0]}}&\\
&\frac{e^{-\!\gamma}\sinh{[u]}}{\sinh{[u_0]}}&1&\\
&&&\frac{\sinh{[u_0\!\pm\! u]}}{\sinh{[u_0]}}
\end{array}
\right),&\\
&\check{R}_{24}^{\pm}(a_0,a_x;u)=\left(\!\!\begin{array}{cccccccc}{r_{12}}_{++}^{++}&&&&&&&\\
&{r_{11}}_{++}^{++}&&&&&{r_{11}}_{-+}^{+-}&\\
&&{r_{11}}_{+-}^{+-}&&&&&\\
&&&{r_{12}}_{+-}^{+-}&{r_{12}}_{-+}^{+-}&&&\\
&&&{r_{12}}_{+-}^{-+}&{r_{12}}_{-+}^{-+}&&&\\
&&&&&{r_{11}}_{-+}^{-+}&&\\
&{r_{11}}_{+-}^{-+}&&&&&{r_{11}}_{+-}^{+-}&\\
&&&&&&&{r_{12}}_{--}^{--}
\end{array}\!\!
\right)&\\
&\check{R}_{42}^{\pm}(a_0,a_x;u)=
=\left(\!\!\begin{array}{cccccccc}{r_{21}}_{++}^{++}&&&&&&&\\
&{r_{21}}_{+-}^{+-}&&&&&{r_{21}}_{-+}^{+-}&\\
&&{r_{11}}_{++}^{++}&&&&&\\
&&&{r_{11}}_{+-}^{+-}&{r_{11}}_{-+}^{+-}&&&\\
&&&{r_{11}}_{+-}^{-+}&{r_{11}}_{-+}^{-+}&&&\\
&&&&&{r_{11}}_{-+}^{-+}&&\\
&{r_{21}}_{+-}^{-+}&&&&&{r_{21}}_{+-}^{+-}&\\
&&&&&&&{r_{21}}_{--}^{--}
\end{array}\!\!
\right)&
\ena
\end{small}}
The matrix elements of the sub-matrices $r_{11}$ just coincide with the elements of matrix (\ref{r22pm}) up to an overall function, the next sub-matrices can differ by constants:
\bea
{r}_{11}(u)=\check{R}_{22}^{\pm}(a_0 u,u_0,\gamma),\quad
{{r}_{21}}(u)=e^{\alpha u}\check{R}_{22}^{\pm}(a_x u,u_x,\beta),\quad
{{r}_{12}}(u)=e^{\alpha' u}\check{R}_{22}^{\pm}(a_x u,u_x,\beta')
\ena
In the equations (\ref{424}) the parameters $a,\;\alpha, \;\beta,\; \gamma$ are independent ones, but the constant parameters $(u_0,\; u_x;\;a_0,\;a_x)$ must be  same for all the cells of $\check{R}_{24/42}^{\pm}$ and coincide with  the constants  in $\check{R}_{22}^{\pm}$, e.g. ($u_x=u_0$, $a_0=1$, $a_x=1$): 
\bea
\check{R}^{\pm}_{22}(u-v)\check{R}_{24}^{\pm}(1,1;u)\check{R}_{22}^{\pm}(v)=
\check{R}_{24}^{\pm}(1,1;v)\check{R}_{22}^{\pm}(u)\check{R}_{24}^{\pm}(1,1;u-v).\ena
But the inhomogeneous GYBEs $(\{2,4,2\},\{2\},\{1\})$ admit the multi-parametric solutions:
\bea\check{R}_{24}^{\pm}(a_0,\! a_x\!;u\!-\!v\!)\check{R}_{42}^{\pm}(a_0,\! a_x\!;\!u)\check{R}_{24}^{\pm}(a_0,\! a_x\!;\!v)\!=\!
\check{R}_{42}^{\pm}(a_0,\! a_x;\!v)\check{R}_{24}^{\pm}(a_0,\! a_x;\!u)\check{R}_{42}^{\pm}(a_0,\! a_x;\!u\!-\!v\!).
\label{24g}\ena
Moreover, the matrices $\check{R}_{42/24}(u)$, which satisfy to only above equations (\ref{24g}), can consist of two different YBE-solutions $\check{r}_{11}(u)$ and $\check{r}_{21/12}(u)$, as these equations are superpositions of two
 independent equations.
 However here the matrices  $\check{R}_{24}$ and $\check{R}_{42}$ are different in forms (and can not be presented as one operator $\check{R}_{222}$ even for the identical parameters), meanwhile for the case of the previous subsection, even the multi-parametric solutions for  two matrices $\check{R}_{24}$ and $\check{R}_{42}$ can coincide one with another, and thus can be considered as solutions to ordinary GYBEs $(2,3,1)$.

The corresponding Hamiltonian operators for $XXZ$-type solutions are the following ones, providing that \\
$ S_{2k+1}^{x,y,z}=\sigma_{2k+1;1}^{x,y,z}+\xi
\sigma_{2k+1;2}^{x,y,z}:$
\bea
H_{24/42}^{+}=J_0\sum_k\left( S_{2k}^{x}\sigma_{2k+1}^x+S_{2k}^{y}\sigma_{2k+1}^y+J_z S_{2k}^{z}\sigma_{2k+1}^z+
\sigma_{2k-1}^x S_{2k}^{x}+\right.\\\nn \sigma_{2k-1}^y S_{2k}^{y}+J_z
\sigma_{2k-1}^z S_{2k}^{z}+
\left. J_{\alpha}[\sigma_{2k-1;1}^z]^2 [\sigma_{2k}^{z}]^2+J_{\alpha'} [\sigma_{2k}^{z}]^2[\sigma_{2k+1;1}^z]^2\right),
\ena
\bea
H_{24/42}^{-}=J_0\sum_k\left(S_{2k}^{x}\sigma_{2k+1}^x+S_{2k}^{y}\sigma_{2k+1}^y+J_z[ S_{2k}^{z}+\sigma_{2k+1}^z]+ \sigma_{2k-1}^x S_{2k}^{x}+\right.\\\nn \sigma_{2k-1}^y S_{2k}^{y}+J_z
[\sigma_{2k-1}^z +S_{2k}^{z}]+\left.
\; J_{\alpha}[\sigma_{2k-1;1}^z]^2 [\sigma_{2k}^{z}]^2+J_{\alpha'} [\sigma_{2k}^{z}]^2[\sigma_{2k+1;1}^z]^2\right),
\ena

The constants $J,\xi$ are easily can be connected with constants of the $R$-matrices. In the last summands we have taken into account the diagonal terms, induced by the terms with $\alpha\; \alpha'$ (here for simplicity we omitted some  constants, as example $\gamma=0$).
The similar inhomogeneous 1D lattice Hamiltonian operators emerged from  the $XYZ$ and other solutions with Jacobi elliptic functions is straightforward.

\paragraph{Statement III} The inhomogeneous GYBEs in the form
\bea\label{gybe123}
\check{R}_{{N}_1{N}_2}(u,v)
\check{R}_{{N}_2{N}_3}(u)\check{R}_{{N}_1{N}_2}(v)=
\check{R}_{{N}_2{N}_3}(v)
\check{R}_{{N}_1{N}_2}(u)\check{R}_{{N}_2{N}_3}(u,v)
\ena
when $\mathcal{N}_2>\mathcal{N}_1 \geq \mathcal{N}_3$ consist of at least $[\mathcal{N}_2-\mathcal{N}_1]$ independent YBE defined on $4\times 4$  sub-matrices $r_{n_i n_j}(u)$, thus they have independent solutions, i.e. the corresponding matrices can contain different eight-vertex kind solutions simultaneously and even with different
spectral parameters. However these matrices can not be used for constructing integrable chains, as for transfer matrices commutations,
together with the equations (\ref{gybe123}),  one must consider also the equations:
\bea\label{gybe231}
\check{R}_{{N}_2{N}_3}(u,v)
\check{R}_{{N}_3{N}_1}(u)\check{R}_{{N}_2{N}_3}(v)=
\check{R}_{{N}_3{N}_1}(v)
\check{R}_{{N}_2{N}_3}(u)\check{R}_{{N}_3{N}_1}(u,v)
\ena
\bea\label{gybe312}
\check{R}_{{N}_3{N}_1}(u,v)
\check{R}_{{N}_1{N}_2}(u)\check{R}_{{N}_3{N}_1}(v)=
\check{R}_{{N}_1{N}_2}(v)
\check{R}_{{N}_3{N}_1}(u)\check{R}_{{N}_1{N}_2}(u,v)
\ena
Consideration of all the equations (\ref{gybe123}, \ref{gybe231}, \ref{gybe312}) brings to similar type YBE-solutions (with the same parameters) for all the sub-matrices.

\paragraph{Note,} that the spin quantities indexed by $p=\{1,...,P\}$ at the site $i$, $\sigma_{i;p}$, can be considered

as the spins situated at the sub-chains with the indexes $\{i_1,i_2,...,i_P\}$. However, taking into account, that there
is no interaction between the spins $\sigma_{i_p}$ with the same $i$, there is preferable to
keep the previous notations $\sigma_{i;p}$, which indicates that at each site $i$ there is a set of $\frac{1}{2}$-spins of number $P$.

\end{document}